%% file: main-document.tex
\documentclass[
authoryear,
preprint,times,review,longtitle,3p
]{elsarticle}
\usepackage[T1]{fontenc}
\usepackage[publisher=elsevier]{arXiv-notices}
\journal{International Journal of Human-Computer Studies}

\usepackage{graphicx}
\usepackage{tabularx}
\usepackage{enumitem}
\usepackage{tikz}
\usepackage{makecell}
\usepackage{multirow}
\usepackage{acronym}
\usetikzlibrary{mindmap,trees,shapes,positioning,arrows.meta}
\usepackage[edges]{forest}
\usetikzlibrary{shadows.blur}
\usepackage[hidelinks]{hyperref}

\usepackage{float}
\usepackage[utf8]{inputenc} % Unicode input
\usepackage[T1]{fontenc}
\usepackage{lmodern}

\usepackage{ltablex}
\keepXColumns
\renewcommand{\arraystretch}{1.2} 
\usepackage{graphicx}
\usepackage{caption}
\usepackage{pdflscape}
\usepackage{booktabs} % \toprule \midrule \bottomrule
\usepackage{array}    % for >{\raggedright\arraybackslash}
\usepackage{ragged2e} % for ragged alignment in cells
\usepackage{longtable} % if needed for very long tables

\makeatletter
\def\paragraph{\secdef{\els@aparagraph}{\els@bparagraph}}
\def\els@aparagraph[#1]#2{\elsparagraph[#1]{#2.}}
\def\els@bparagraph#1{\elsparagraph*{#1.}}

\renewcommand\elsparagraph{\@startsection{paragraph}{4}{\parindent}%
           {0pt}%
           {-6\p@}%
           {\normalfont\itshape}}
\makeatother

\usepackage[horizontal,grid=lightgray,skipempty]{credits}

\acrodef{HCI}{Human-Computer Interaction}
\acrodef{HCAI}{Human-Centered Artificial Intelligence}
\acrodef{AI}{Artificial Intelligence}
\acrodef{CCT}{Code Completion Tool}
\acrodef{IDE}{Integrated Development Environment}

\begin{document}

\begin{frontmatter}

\title{Bug Detective and Quality Coach: Developers’ Mental Models of AI-Assisted IDE Tools}

\author[uniba]{Paolo Buono}
\ead{paolo.buono@uniba.it}
\ead[orcid]{https://orcid.org/0000-0002-1421-3686}

\author[unisa]{Mary Cerullo}
\ead{mcerullo@unisa.it}
\ead[orcid]{https://orcid.org/0009-0004-5880-6717}

\author[unisa]{Stefano Cirillo}
\ead{scirillo@unisa.it}
\ead[orcid]{https://orcid.org/0000-0003-0201-2753}

\author[uniba]{Giuseppe Desolda\corref{cor1}}
\ead{giuseppe.desolda@uniba.it}
\ead[orcid]{https://orcid.org/0000-0001-9894-2116}
\cortext[cor1]{Corresponding author}

\author[uniba]{Francesco Greco}
\ead{francesco.greco@uniba.it}
\ead[orcid]{https://orcid.org/0000-0003-2730-7697}

\author[unimol]{Emanuela Guglielmi}
\ead{emanuela.guglielmi@unimol.it }
\ead[orcid]{https://orcid.org/0000-0002-5443-1303
}

\author[unisa]{Grazia Margarella}
\ead{gmargarella@unisa.it}
\ead[orcid]{https://orcid.org/0009-0007-5520-2872}

\author[unisa]{Giuseppe Polese}
\ead{gpolese@unisa.it}
\ead[orcid]{https://orcid.org/0000-0002-8496-2658}

\author[unimol]{Simone Scalabrino}
\ead{simone.scalabrino@unimol.it }
\ead[orcid]{https://orcid.org/0000-0003-1764-9685}

\author[uniba]{Cesare Tucci}
\ead{cesare.tucci@uniba.it}
\ead[orcid]{https://orcid.org/0000-0001-5181-7115}

\affiliation[uniba]{%
    organization={Department of Computer Science, University of Bari Aldo Moro},
    addressline={Via E. Orabona 4},
    city={Bari},
    postcode={70125},
    country={Italy}
}
\affiliation[unisa]{%
    organization={University of Salerno},
    addressline={Via Giovanni Paolo II 132},
    city={Fisciano (Salerno)},
    postcode={84084},
    country={Italy}
}

\affiliation[unimol]{%
    organization={University of Molise},
    addressline={Via Duca degli Abruzzi, 67},
    city={Termoli (Campobasso)},
    postcode={86039},
    country={Italy}
}

% --- Autori/affiliazioni (commenta o compila) ---
% \author[inst1]{First Author}
% \author[inst1]{Second Author}
% \affiliation[inst1]{organization={Department/University},
%   city={City}, country={Country}}

\begin{abstract}
AI-assisted tools support developers in performing cognitively demanding tasks such as bug detection and code readability assessment. Despite the advancements in the technical characteristics of these tools, little is known about how developers mentally model them and how mismatches affect trust, control, and adoption. We conducted six co-design workshops with 58 developers to elicit their mental models about AI-assisted bug detection and readability features. It emerged that developers conceive bug detection tools as \textit{bug detectives}, which warn users only in case of critical issues, guaranteeing transparency, actionable feedback, and confidence cues. Readability assessment tools, on the other hand, are envisioned as \textit{quality coaches}, which provide contextual, personalized, and progressive guidance. Trust, in both tasks, depends on the clarity of explanations, timing, and user control. A set of design principles for Human-Centered AI in IDEs has been distilled, aiming to balance disruption with support, conciseness with depth, and automation with human agency.
\end{abstract}

\begin{keyword}
Human–AI collaboration\sep
AI-assisted IDEs\sep bug detection\sep code readability\sep mental models\sep
trust
\end{keyword}

\end{frontmatter}

% ===== Teaser: convertita in figura a tutta pagina/colonna =====
\begin{figure*}[t]
  \centering
  \includegraphics[width=\textwidth,trim=0 50 0 70,clip]{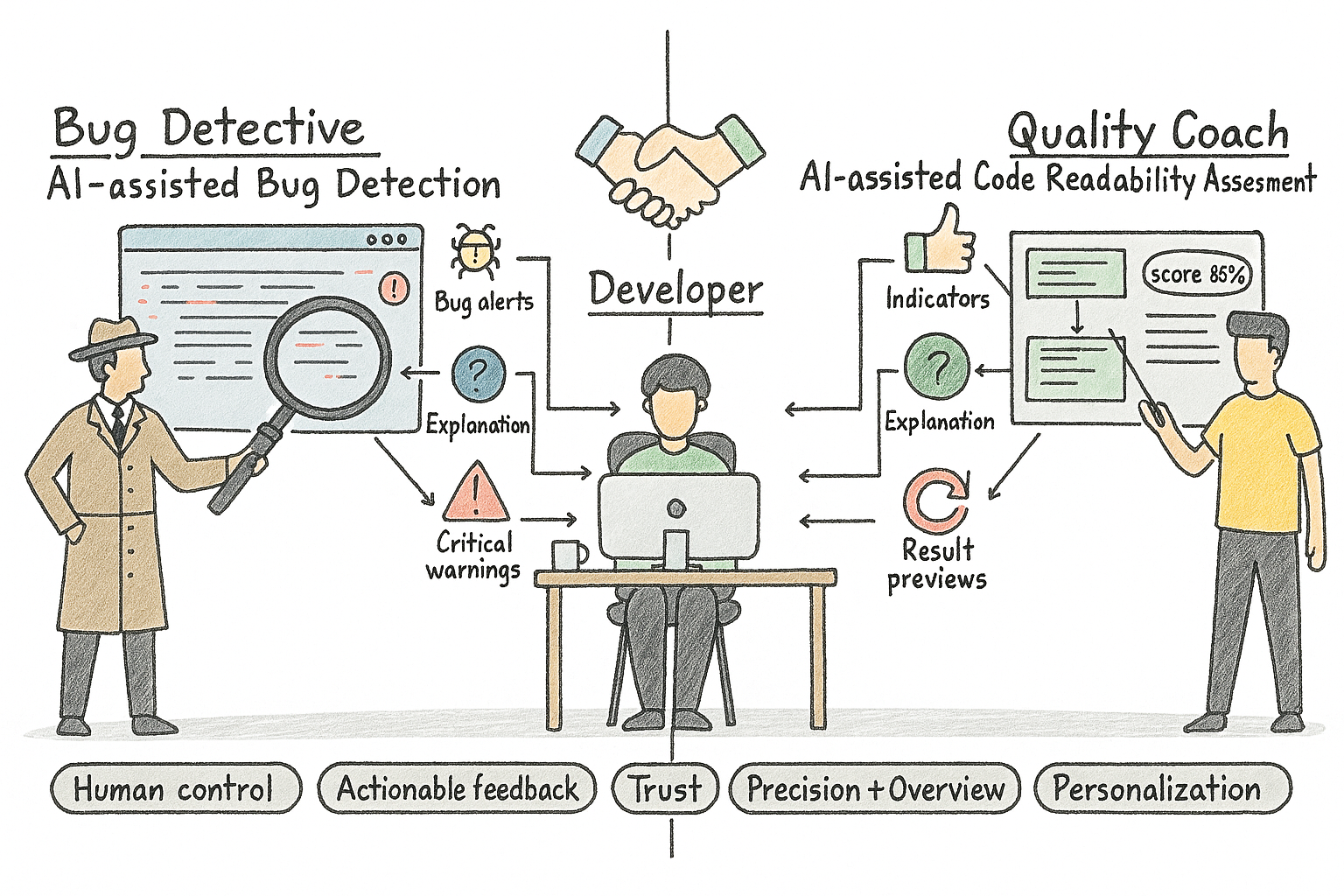}
  \caption{Metaphorical Representation of AI-Assisted IDE Mental Models. It illustrates the two primary mental models identified for AI-assisted tools: the "Bug Detective" (left) and the "Quality Coach" (right). It highlights how these "partners" collaborate with the developer to offer actionable feedback, critical warnings, and personalized guidance, emphasizing the crucial role of "Trust" fostered through explanation clarity, user control, and adaptive timing.}
  \label{fig:mental_models}
\end{figure*}

\section{Introduction}
\texorpdfstring{\acf{AI}}{AI} has a significant impact on how people interact with their tools. This transformation is particularly evident in software development, since modern \acfp{IDE} increasingly adopt AI-enhanced tools, transforming how developers write, review, and maintain code. Code writing is among the first development tasks that benefited from this evolution \cite{Weber2024Significant,  Cui2024Effects, Peng2023Impact}. Thanks to the improvement of AI models, more complex, cognitive-intensive activities such as \textit{bug detection} \cite{Li2024Static, Wen2024Automatically, Allamanis2021SelfSupervised} and \textit{code readability assessment} \cite{vitale2023using} can be assisted. Aspects such as software quality, security, and long-term maintainability can be improved by automating these tasks. 

To implement effective human-AI collaboration, integrating AI solutions in IDEs should extend beyond technical aspects (e.g., algorithms or AI models \cite{izadi2024language, husein2024large}). In fact, a flawed design of such tools — failing to consider developers’ mental models — can pose risks such as deskilling \citep{Sambasivan2022Deskilling}, insecure code \citep{Sergeyuk2025Using}, and friction when AI outputs conflict with developers’ judgments. The \acf{HCAI} paradigm advocates that AI systems, such as those integrated inside IDEs, should be designed and evaluated with user involvement to improve task performance and satisfaction \citep{Shneiderman2022HumanCentered, Xu2019HumanCentered, Desolda2024Humancentred}. 

Despite these advances, research has mainly focused on improving the accuracy and performance of AI-assisted tools (e.g., Li et al., 2024; Wen et al., 2024). However, from an HCI perspective, less attention has been devoted to understanding how developers conceptualize, interpret, and trust these AI-based tools. This gap limits our understanding of how AI integration impacts developers’ mental models, interaction patterns, and sense of agency. Research on human-centered AI (e.g., \cite{Amershi2019Guidelines, shneiderman2020human, liao2020questioning}) highlights that a successful human–AI collaboration depends not only on algorithm accuracy but also on users’ mental models, the design of explanations, and the process of building trust. Still, in the context of IDEs, these factors have not been thoroughly studied. 

Some HCI researchers have already begun investigating these tools, primarily focusing on developers' preferences and expectations for code completion tasks (e.g., \cite{muaruașoiu2015empirical, Liang2024LargeScale, Xu2022IDE, Bird2022Taking, Zhang2023Demystifying}). However, little is known about how developers interact with AI for quality-centric tasks such as bug detection and code readability assessment. These two tasks are closely related, as they require higher-order reasoning processes (for instance, evaluating impact, understanding context, and anticipating consequences), and both play a central role in shaping developers’ trust in AI-based tools. Importantly, they are often interdependent: poor readability can determine a more complex bug detection process, while identifying bugs may highlight readability issues that need to be addressed. Studying them together enables us to identify shared challenges in interaction design (e.g., the need for precise, contextualized, and adaptable system feedback), while also capturing potential synergies between them. Furthermore, by analyzing two distinct but comparable tasks, we can derive design implications that are more generalizable beyond a single use case. 

Indications and suggestions not properly reported and explained by the system can waste developers' time, introduce errors, and erode trust for these tasks. The consequence of this misalignment between developers' mental models and these tools emerged, for example, in a recent study stating that around 40\% of AI-generated suggestions are accepted by developers on first use, mainly due to the limited control of the suggestion, missing or insufficient details of the explanations, and poor interaction design \cite{Liang2024LargeScale}. Other studies have also reported that developers need finer-grained suggestions, trustworthy explanations \cite{Sergeyuk2025Using, Wang2023How}, as well as flexibility in interaction modalities (e.g., inline vs. sidebar layouts \cite{Vaithilingam2023IntelliCode, muaruașoiu2015empirical}).

Building on this gap, this study investigates how developers interact with AI-assisted IDEs, specifically for the tasks of \textit{bug detection} and \textit{code readability assessment}. To this aim, our study aims to answer the following three research questions:
\begin{enumerate}[label=RQ\arabic*)]
 \item How do developers mentally model their interactions? 
 \item What do developers expect regarding explanations, personalization, and control?
 \item What design principles match AI tools with developers’ mental models?
 \end{enumerate}

Given the exploratory and qualitative nature of the problem, we conducted an elicitation study with 58 developers across six co-design workshops to answer these research questions. Our contribution is threefold:
\begin{enumerate}
    \item \emph{Empirical}: To the best of our knowledge, we conducted the first study that explores the mental models of developers using AI-assisted tools for bug detection and readability assessment.
    \item \emph{Conceptual}: We found that developers see bug detection tools as \textit{bug detectives} and readability assessment tools as \textit{quality coaches}. We also identified tensions between trust, control, and explainability.
    \item \emph{Design}: We identified seven design principles for both tasks, which connect AI outputs with developers’ mental models.
\end{enumerate}

Besides these contributions, this study also informs the design of HCAI systems that promote trust, transparency, and collaboration between humans and AI. Indeed, our contributions extend to a broader discussion on HCAI aspects by addressing key questions such as: How do users trust AI systems that evaluate quality? How should AI systems provide explanations that lower cognitive load? How can we foster a symbiosis between humans and AI? These questions extend beyond IDEs and impact the creation of AI-assisted tools in various professional fields.

\section{Background \& Related Work}
Our study builds on two main areas. The first involves mental models in HCI, which explain how users form expectations of complex systems and why mismatches can hurt trust and usability. The second area relates to AI-assisted developer tools. Research has primarily focused on code completion, but questions remain about more complex tasks, such as bug detection and readability assessment. Combining these perspectives informs our investigation and underscores the importance of human-centered design for AI features in IDEs.

\subsection{Mental Models}

HCI researchers study mental models to explain how users understand system functionality, and to help users predict system behavior and shape their interactions. In this paper, we adopt the classic HCI perspective on mental models, defined as “the internal representations that people form of a system’s structure and processes, which enable them to predict and reason about its behavior” \cite{payne2007mental, johnson1972reasoning, Norman1983Observations}. Mental models are often partial, unstable, and influenced by users’ prior experiences and interaction context. In our study, we treat developers’ mental models as the implicit conceptualizations of how AI-assisted IDE tools operate, revealed through their verbal reasoning and sketches during co-design workshops. These models inform how developers expect the AI to behave, when to trust its suggestions, and how to interpret its explanations.

In recent years, the complexity and lack of transparency of AI systems, such as generative AI, have made it difficult to align mental models with the AI systems \cite{fischer1991importance,fischer2001user,fischer2017exploring, canas2001role,johnson2010mental,norman1986cognitive}. When gaps exist between user expectations and system functionality, it can lead to errors, reduced trust, and inefficient usage \citep{190/sym130.33050795, yarosh2017locked}. 

This issue is especially noticeable in AI-assisted developer tools. Developers often struggle to understand the source or reliability of the suggestions they receive. In a recent study by Wang et al., it was found that not only developers claim high-performing tools, but they also require predictability and clarity \cite{Wang2023How}. 

The relevance of mental models in AI-based IDE tools emerges along four dimensions. First, they support \textit{interaction} by helping developers judge the usefulness of suggestions and maintain coding flow. Second, they foster \textit{trust}, since transparency and interpretability are necessary to overcome the black-box nature of such tools \citep{Desolda2024Humancentred}. Third, they might help \textit{reduce cognitive workload} by mitigating interruptions and enhancing productivity, whereas poor alignment increases mental burden. Finally, they provide \textit{learning support}, especially for novice developers, by suggesting improvements, alerting to bugs, and strengthening coding competence. Therefore, designing these tools in accordance with users’ mental models enhances usability and adoption \citep{Sergeyuk2025Using}, a point underscored by the finding that 54\% of the developers explicitly call for better AI-based IDE tools \cite{Wang2023How}.

Investigating mental models, however, remains challenging due to their abstract, incomplete, and often subconscious nature. For these reasons, it is essential to consider adequate methodologies to elicit users' mental models, such as design workshops \citep{Jacko2012Human}, interviews \citep{Rogers2023Interaction}, think-aloud protocols \citep{Lewis1982Using}, and scenario-based design \citep{Rosson2012ScenarioBased}. Insights obtained through these solutions can inform design strategies and strengthen \ac{HCI} research in AI-augmented coding environments. 

The present study employs a co-design workshop to elicit developers’ mental models, following previous work \cite{desolda2025understandingusermentalmodels}, while extending the focus to bug detection and code readability assessment. The following section details the study design, by focusing on two Software Engineering tasks: \textit{bug detection}/\textit{defect prediction} and \textit{code readability assessment}, and by reporting previous attempts to (even partially) define mental models for them.

\subsection{Bug Detection in HCI}
Bug detection tools are implemented in IDEs to automatically identify errors or defect-prone code segments that may cause incorrect behavior, failures, or vulnerabilities in software systems. Aljedaani \textit{et al.} \cite{aljedaani2024empirical} showed that the effectiveness of bug detection tools depends not only on detection accuracy but also on how their warnings and feedback are communicated and integrated into developer workflows. Other studies have demonstrated that developers frequently struggle with unclear warning messages, as current tools often fail to provide answers and explanations that align with users' reasoning, leading to misinterpretation and a loss of trust and efficiency \cite{10.1145/2970276.2970347, smith2015questions}. This is particularly critical in security-focused Static Analysis Tools (SATs), where notifications often fail to enable developers to act without manually checking \cite{10.1145/3411764.3445616}. To address this, previous works have tried to design tools for generating explanations aligned with developers' self-explanatory processes \cite{barik2016static,10172643,10.1145/3613904.3642239}. 
For instance, ASIDE has been conceived as a user-centered integration of SATs as developer IDE plugins, which provides multi-level and contextualized explanations when identifying vulnerable snippets of code, increasing developers' accuracy and awareness \cite{thomas2016questions}. However, customization is only achieved through code annotations and only affects the content of explanations (not the timing, for example).
Do \textit{et al.} \cite{9124719} recommend a user-centered approach to SATs, emphasizing the integration of developer knowledge, motivations, and usage context to improve usability and effectiveness. 
Ami \textit{et al.} \cite{ami2024false} report that some of the most recurrent concerns industrial developers have regarding the use of SATs are about the trust in the tool's effectiveness and the possibility of manually interacting with the vulnerable section. 
Finally, in a recent empirical study at Microsoft, Christakis and Bird \cite{10.1145/2970276.2970347} found that unclear warnings, slow performance, and workflow disruptions hinder the use of such tools. Developers value tools that highlight best practice violations, align with their workflow, minimize distractions, and guarantee low false positive rates.

While previous work clearly shows that integrating existing bug detection approaches and tools into developers' workflows has several shortcomings, the literature still lacks a comprehensive understanding of developers' mental models regarding such approaches.

\subsection{Code Readability Assessment in HCI}
Code readability is acknowledged as a key factor in software quality. Readable code facilitates understanding, debugging, and maintenance activities \cite{dantas2021readability}. The \textit{code readability assessment} tools automatically evaluate code portions (snippets, methods, or classes) regarding such an aspect. Most existing code readability assessment approaches work by categorizing a given code element as \textit{readable} or \textit{unreadable} \cite{buse2008metric,buse2009learning,posnett2011simpler,dorn2012general,scalabrino2016improving,scalabrino2018comprehensive}, while other approaches also use a third category (\textit{neutral} or similar) \cite{mi2018improving,mi2022towards,mi2023graph,vitale2025personalized}.
Previous work has shown that one of the main limitations of existing code readability assessment approaches is that they are too generic and not sufficiently tailored to the developer or project at hand. Vitale et al. \cite{vitale2025personalized} explored the possibility of evaluating readability on an individual basis. However, their study showed that personalized few-shot models based on LLM performed worse than generic feature-based models, highlighting the difficulty of modeling developers' specific perceptions and the need for more robust datasets.

Past works also overlook the questions of usability (e.g., how readability checkers fit in developers’ workflows) and explainability (e.g., providing a clear rationale with a readability score). Investigations on support for user control or collaborative customization are also lacking. The misalignment between algorithmic models and developers’ mental models is supported in empirical findings: Sergeyuk et al. \cite{sergeyuk2024assessing} found only weak correlations between a tool’s readability evaluation and developers’ own judgments on the clarity of the code. This can lead to increased cognitive fatigue (developers must double-check or modify tool suggestions) and decreased trust in the AI assistant. Al Madi \cite{almadi2022how} also observed a trust risk, as when the AI produces code that looks highly readable, developers may become over-reliant and inspect it less carefully. 

While the literature has laid a technical foundation for evaluating code readability, it has not yet addressed the human-centered aspects aligning with developer mental models and custom readability criteria. Our work aims to fill this gap by investigating how developers prefer and would engage with an AI-assisted readability tool, thereby informing the design of systems that provide readability guidance aligned with developers’ needs and expectations.

\section{Methodology}

To investigate developers' mental models of AI-assisted IDE features, we conducted an elicitation study using co-design workshops as our primary data collection method. In these workshops, groups of developers engaged with realistic scenarios involving two AI-driven tasks: \textit{automated bug detection} and a \textit{code readability assessment}. They discussed their understanding and expectations of these tools, both with and without explanations. 

\subsection{Study Design}

This study employs a workshop-based elicitation methodology instead of interviews or questionnaires because presenting concrete and realistic situations helps participants express their expectations and preferences in specific contexts~\cite{carroll2003making, rosson2002usability}. Moreover, the adopted methodology is well-suited for investigating complex systems such as AI-powered IDE tools, where abstract concepts must be grounded in practical usage scenarios. 

This study was conducted over six sessions of co-design workshops, which consisted of structured discussions guided by specific questions. This approach enabled a systematic exploration of user mental models across various scenarios, ensuring consistent data collection. The workshop format encouraged group interactions, allowing participants to build on each other’s ideas and perspectives and point out insights that might not have emerged in individual interviews.

To improve the quality of the mental model's elicitation and the analysis and reporting of the results, we adopted the 5W model \cite{Lasswell1974Structure}. This model is used in domains like journalism, customer analysis, and more generally in problem-solving, to describe and analyze the complete story of a fact. In HCI, it might be employed for mental model elicitation and user analysis \citep{Jacko2012Human,Desolda2017Empowering,Duric2002Integrating}. The complete list of the questions used in the study is reported in Table \ref{tab:questions}.

\subsection{Participants}
\label{sec:participants}

We recruited 58 participants from three universities in Southern Italy, i.e., University of Bari, University of Salerno and University of Molise. The mean age was 25.2 years (SD $\approx$ 7.5). Of 58 participants, 43 identified as male and 15 as female. 
A majority (35) of the participants stated not to have obtained the Bachelor's degree yet; 5 had it already, 9 had a Master's degree, and the remaining 9 were PhDs.

%Participants had a mean programming experience of 5.9 years (SD = 6.5).
The sample included both students (40) and professional developers (18) with varying levels of programming experience (mean = 5.9 years of experience; SD $\approx$ 6.5).

Participants' backgrounds spanned various programming languages, including Java, Python, C, C++, PHP, JavaScript, and HTML. Some participants also confirmed they had worked with Visual Basic, Go, Ruby, Swift, and Kotlin. This variety showcases a wide range of coding practices and skills, providing a great starting point for exploring how AI-assisted IDE tools function in various contexts.

No prior experience with AI-assisted development tools was required; yet, 42 participants reported using tools such as GitHub Copilot or AI-based code completion systems.  Some others (13) did not have prior usage experience but claimed to \textit{"know what they are and how they may work"}. Only 3 participants had no previous knowledge of such tools.

All relevant ethical guidelines were followed, and the study was approved by the University of Bari's Ethical Review Board. 

\subsection{Procedure}
One month before the start of the workshops, each of the three universities sent around 50 emails (for a total of more than 150 emails) to PhD, master's, and bachelor's students who had completed at least three programming courses in their academic program, as well as to employees of IT companies, inviting them to participate in the study. A total of 58 volunteers agreed (approximately 20 from each university) and completed a questionnaire requesting their availability for the study week, allowing them to schedule their participation according to their preferences (The questionnaire is available in the Appendix, Section \ref{appendix-questionnaire}.

A total of 6 \ac{HCI} researchers (2 from each university) conducted the study, following the same protocols. One researcher served as the conductor, and the other acted as the observer. Each university research group planned the workshops on a single day in a quiet university laboratory. Each workshop was planned to last around 120 minutes. 

A total of six groups of participants were created (2 groups per University). No group exceeded 10 participants. Each session began with a brief introduction (approximately 5 minutes), during which participants were informed about the confidentiality and voluntary nature of the study, as well as its goals. All the participants agreed. Then, participants sat around a table and were provided with pens and paper to sketch preliminary ideas and solutions when needed. 

The facilitator then presented the first scenario related to the first task (among bug detection or code readability assessment). The presentation order of the two tasks was balanced across groups to control for order effects. The order of the two scenarios for each task was maintained in sequence, as the second scenario depends on the first one. 

For the task of bug detection, participants discussed two scenarios: \textit{Bug Detection – Basic}, which involved addressing how AI-driven alerts should be designed to support developers' workflows, and \textit{Bug Detection – Explanations}, which focused on integrating AI-generated explanations into developers' workflows. Participants were guided through specific questions, which are reported in Table \ref{tab:questions}. For the task of \textit{Code readability measurement}, participants discussed two scenarios: one addressing the metrics and visual presentation of readability indicators, and one focusing on explanations accompanying readability scores. For each task scenario, participants discussed specific questions reported in Table \ref{tab:questions}. The text of all the scenarios is attached in the Appendix (Section \ref{appendix-scenarios}). 

% ...
\newcolumntype{L}{>{\RaggedRight\arraybackslash}p{9.5cm}}

{\small\setlength{\tabcolsep}{5pt}
\begin{longtable}{|c|c|c|L|c|}
\caption{Full list of workshop questions across tasks and scenarios, with corresponding research questions addressed.}
\label{tab:questions}\\
\hline
\textbf{Task} & \textbf{Scen.} & \textbf{QID} & \textbf{Question} & \textbf{RQs} \\
\hline
\endfirsthead
\multicolumn{5}{c}{\tablename\ \thetable{} — continued}\\
\hline
\textbf{Task} & \textbf{Scen.} & \textbf{QID} & \textbf{Question} & \textbf{RQs} \\
\hline
\endhead
\hline
\multicolumn{5}{r}{\footnotesize\itshape Continued on next page}\\
\endfoot
\hline
\endlastfoot
&  & Q1 & What technical details (e.g., error type, affected code segments) and contextual cues should be included in the bug alert to support a developer’s decision-making? & 1,2 \\
\cline{3-5}
& \multirow[c]{5}{*}{1} & Q2 & Which visual design elements (e.g., color coding, icons, animation) should be used to display the bug alert effectively? & 1 \\
\cline{3-5}
\multirow{10}{*}{\rotatebox[origin=c]{90}{\textbf{(A)} Bug Detection}} & & Q3 & Where in the IDE should the bug alert be displayed, and how can its placement adapt dynamically to the user’s current focus to minimize disruption? & 1,3 \\
\cline{3-5}
& & Q4 & When in the development process should the bug alert be activated? & 1 \\
\cline{3-5}
& & Q5 & How can the bug alert be customized to cater to varying levels of developer expertise and different coding contexts? & 2,3 \\
\cline{2-5}
& & Q6 & What key aspects of the bug (e.g., underlying cause, potential impact, suggested fixes) should be explained to support deeper understanding? & 2 \\
\cline{3-5}
& \multirow[c]{5}{*}{2} & Q7 & How should the explanation be presented (e.g., inline tooltip, expandable sidebar) to balance clarity with unobtrusiveness? & 2,3 \\
\cline{3-5}
& & Q8 & Where in the IDE should the explanation be displayed to ensure visibility without interrupting the workflow? & 1,3 \\
\cline{3-5}
& & Q9 & When should the explanation be shown—automatically as part of the alert or on-demand? & 1,2 \\
\cline{3-5}
& & Q10 & How can the explanation be customized based on the developer’s expertise and context? & 2,3 \\
\hline
& & Q11 & What specific metrics and feedback (e.g., readability score, contributing factors such as naming conventions and code structure) should be reported? & 1,2 \\
\cline{3-5}
\multirow{10}{*}{\rotatebox[origin=c]{90}{\textbf{(B)} Code Readability Assessment}} & \multirow[c]{5}{*}{1} & Q12 & How should readability indicators be visually presented (e.g., charts, color-coded indicators) within the IDE? & 1,3 \\
\cline{3-5}
& & Q13 & Where should the code readability feature be shown in the UI to minimize disruption? & 1,3 \\
\cline{3-5}
& & Q14 & When during the coding process should readability indicators be activated? & 1 \\
\cline{3-5}
& & Q15 & How can readability indicators be customized to suit different experience levels and project contexts? & 2,3 \\
\cline{2-5}
& & Q16 & What factors contributing to the readability score should be explained? & 2 \\
\cline{3-5}
& \multirow[c]{5}{*}{2} & Q17 & How should the explanation be presented (e.g., inline comment, pop-up, expandable section)? & 2,3 \\
\cline{3-5}
& & Q18 & Where in the IDE should the explanation be located? & 1,3 \\
\cline{3-5}
& & Q19 & When is the optimal time to present the explanation—automatically or only when requested? & 1,2 \\
\cline{3-5}
& & Q20 & How can the explanation be tailored (e.g., technical depth, language complexity) to different developer profiles? & 2,3 \\
\end{longtable}
}

Each scenario presentation was followed by a structured elicitation phase that lasted approximately 20 minutes. For each question, participants were asked to think individually for approximately one minute and then engage in a collective discussion to refine their ideas progressively. At the end of each task, a 5-minute debriefing was held to review the most critical points. At the conclusion of the first task, participants took a 10-minute break. Then, the second task started following the same procedure as the first. At the end of the second task, each participant was asked to complete a post-test questionnaire to collect their demographic and skill information (reported in Section \ref{sec:participants}), as well as any additional comments (optional).

\subsection{Data Collection and Analysis}

Different data were collected during the six workshops, i.e., 1) the notes taken by the researchers, 2) the audio recordings, and 3) the sketches drawn by participants. Before the thematic analysis, two researchers independently reviewed approximately 80\% of these materials to ensure the accuracy and completeness of the transcriptions and the consistency between notes, recordings, and sketches. Minor discrepancies (e.g., wording or labeling inconsistencies) were discussed and resolved collaboratively before proceeding with the reflexive thematic analysis.

Starting from this data, we conducted a reflexive thematic analysis following Braun and Clarke’s six-phase approach \cite{Braun2006Using}. Two researchers (each with ~15 years of experience in HCI and software engineering) independently familiarized themselves with the full workshop transcripts, notes, and sketches. Familiarization consisted of different readings and memoing; initial codes were then generated at the semantic level and recorded collaboratively in Google Sheets (columns for code label, supporting excerpt, speaker/group, and analytic memos). Coding was an iterative process: the coders met weekly to compare code lists, discuss divergent interpretations, and refine a shared codebook. The discrepancies that emerged were solved through discussion between the two researchers until a consensus was reached. We deliberately did not calculate intercoder reliability metrics, consistent with a reflexive approach that treats coding as a situated, interpretive activity rather than a purely mechanical task. Codes were organized into candidate themes, which were reviewed against the data in successive rounds and refined into final themes that map onto the 5W elicitation framework used during the workshops. We maintained an audit trail (versioned codebook, meeting notes, and analytic memos) and provided an excerpt of the codebook and themes in the appendix (Section \ref{appendix-codebook}) to support transparency.

\section{Results}

This section outlines the themes developed during the thematic analysis, which are linked to the questions that guided the workshop discussions. All the themes, codes, and frequencies are reported as additional material. The names of each theme are presented in italics in the following subsections. Participants' answers are given in double quotes to exemplify how one of the answers led to the identification of a theme. We also specified the number of groups that proposed the solution linked to each theme. These results provide a concrete description of users' thoughts about desirable interaction. 

\subsection{Interaction with Bug Detection Alerts (Task~A, Scenario~1)}

\begin{figure}[ht]
    \centering
    \includegraphics[width=\linewidth, trim=0 4cm 0 0, clip]{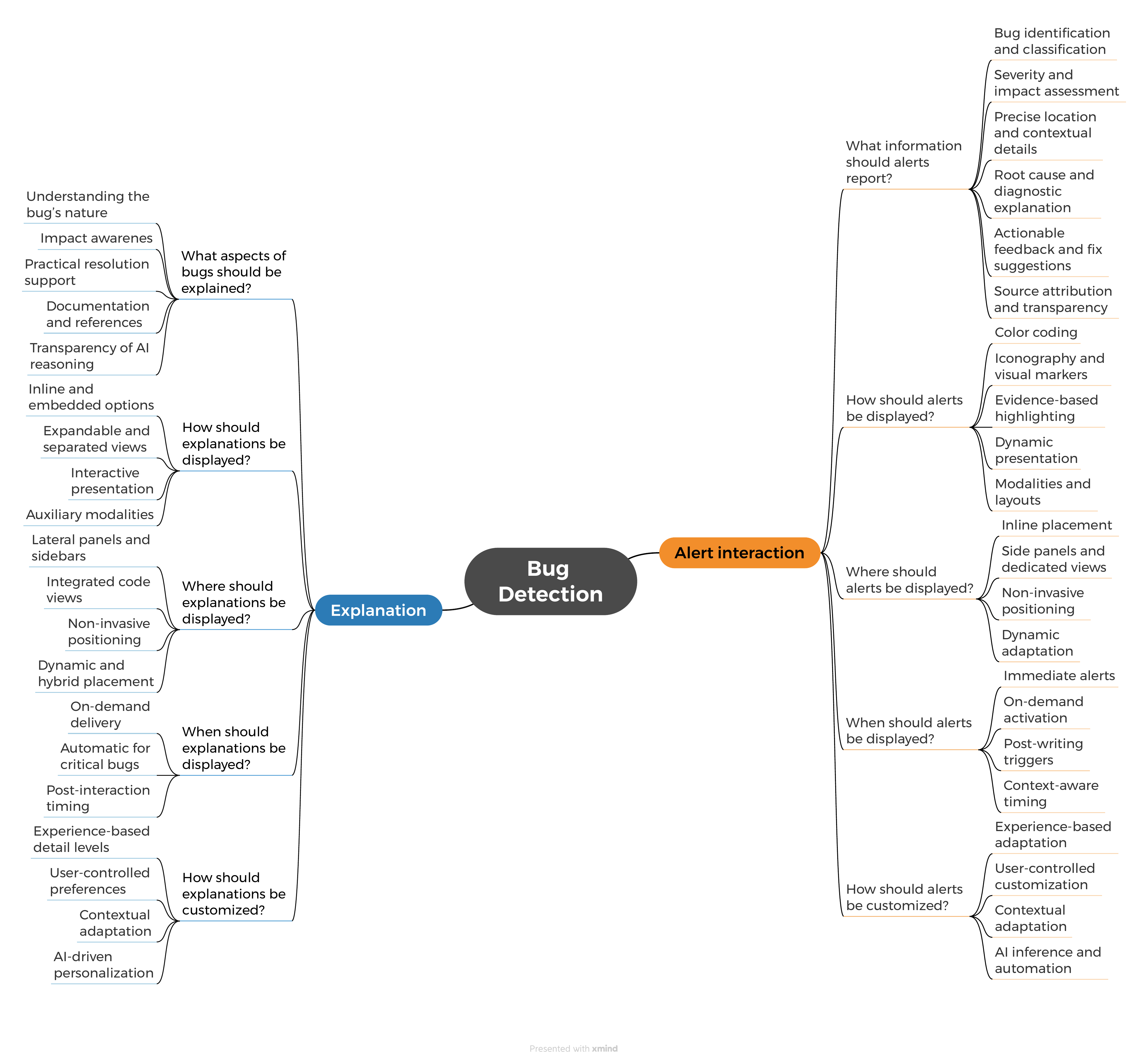}
    \caption{Themes emerged by analyzing the developers' answers to the questions on the bug detection task.}
    \label{fig:findings-ui}
        
\end{figure}

The following findings describe the expectations of participants about \emph{when} and \emph{how} bug alerts should be triggered and displayed, \emph{where} they should appear in the IDE, \emph{what} information they must contain to support the selection and action, and \emph{how} the alert experience should be customized to different users and contexts. Figure \ref{fig:findings-ui} depicts an overall view of the results.

\paragraph{What information should alerts report?}
Participants consistently requested concise, immediate content that enables rapid assessment and confident action. In particular, bug alerts should report:
\begin{itemize}
  \item \emph{Bug identification and classification} (5 groups): Outputting the error class/type or name and, where applicable, the relevant rule or detector allows to set accurate expectations about the nature of the issue \textit{(``I'd like to know the error code, so that I can look for similar cases on the internet''}, - G4).
  \item \emph{Severity and impact assessment} (6 groups): Alerts should clearly report the consequences of the error, especially in the cases of crash risk, security exposure, or performance regression, in order to help developers prioritize issues in critical situations such as when under time pressure \textit{(``What are the risks the bug exposes?''} - G2).
  \item \emph{Precise location and contextual details} (6 groups): Provide file, lines, call site or frame, and salient variables to enrich the alert with observable evidence which is easy to retrieve (\textit{``The alert could include a snapshot representing the system's state at the moment the error occurs.''} - G1).
  \item \emph{Root cause and diagnostic explanation} (5 groups): AI tools should provide a brief ``why'' statement to summarize the main causal mechanism that brought to the error (e.g., untrusted input data, one-by-one in a loop bound), while however saving deeper traces for later use (\textit{``...the impact the bug has on the overall code''} - G3).
  \item \emph{Actionable feedback and fix suggestions} (4 groups): Giving one or two specific next steps (like a targeted edit, a test to run, or a document to read) is preferred, rather than providing generic advice. The system should not perform auto patching without a good reason (\textit{``I'd like the agent to motivate a possible fix that I can either accept or not''}, G1).
  \item \emph{Source attribution and transparency} (3 groups): It is essential to let the users know what the underlying signal is (static rule, learned model, test failure) and give them a rough idea of the level of confidence or uncertainty of predictions/suggestions (\textit{``It would be helpful to know the degree of confidence of the AI in each different case''}, - G5).
\end{itemize}

\paragraph{How should alerts be displayed?}
Developers want alerts that are easy to see and have evidence behind them, as well as visual elements that do not increase cognitive load. According to their thought, alerts could be shown through:
\begin{itemize}
  \item \emph{Color coding} (6 groups): A clear and consistent palette communicates severity and status immediately, while keeping the textual payload minimal (\textit{``...different colors depending on the gravity of the error''}, - G3 and G6). Gentle contrasts are preferred to avoid visual noise when many alerts are displayed simultaneously.
  \item \emph{Iconography and visual markers} (5 groups): The use of familiar symbols (e.g., those common for warnings or info) and compact badges make it easier to scan and help quickly sort through multiple issues, and at the same time saving space on the editor view which could otherwise become too crowded (\textit{``...a danger icon''} - G1).
  \item \emph{Evidence-based highlighting} (5 groups): Alerts should be linked to inline highlights, and gutter markers should anchor each alert to the exact implicated spans (\textit{``...include a link to the lines that cause the bug to localize them quickly''} - G2). This way, the evidence is easily retrievable without requiring manual navigation.
  \item \emph{Dynamic presentation} (3 groups): Animations can signal new findings to give feedback in real-time in a non-obtrusive manner (\textit{``...an animated or flashing icon''} - G1), but a persistent motion of such elements is discouraged by participants. Short walkthroughs are considered acceptable when the issue is complex or important for safety (\textit{``...an animated, short video''} - G5).
  \item \emph{Modalities and layouts} (4 groups): Lightweight, non-blocking features may link to richer detail views (e.g., panel, popover, chatbot) when needed, such that users don't have to deal with modal dialogs that can interrupt the editing of the code (\textit{``...I'd like a pop-up, and a chatbot for optionally explore the bug further''}, - G2).
\end{itemize}

\paragraph{Where should alerts be displayed?}
Placement must be scalable from one issue to many, so that context switching is minimized, and that evidence is kept close to the code:
\begin{itemize}
  \item \emph{Inline placement} (5 groups): Without losing sight of the surrounding code, markers and inline annotations at the exact position of the issue facilitate quick fixes and quick local assessment (\textit{``...right near the code''} - G4).
  \item \emph{Side panels and dedicated views} (5 groups): Docked and expandable panels, or separate tabs, can be used to aggregate all alerts, provide support for sorting and filtering, display history, and disclose features (such as traces, variables, and links) that go beyond what can be displayed inline (\textit{``...a side panel where errors are linked, maybe with the possibility of sorting per gravity''} - G1).
  \item \emph{Non-invasive positioning} (4 groups): Alerts must never block run controls, completions, or selections whenever they are displayed. When space is limited, visual components should prioritize the editor and avoid overlapping essential user interface elements (\textit{``The pop-up that overlays with the code may result in discomfort''} - G2).
  \item \emph{Dynamic adaptation} (4 groups): Especially when the editor is crowded, the system should move from inline alerting to panel display, or it should pin the alert near the active cursor to decrease eye travel (\textit{``...adaptable depending on the coding activity and IDE''} - G3). Considering the viewport state and focus, the system could modify placement accordingly.
\end{itemize}

\paragraph{When should alerts be displayed?}
Participant consensus led to the design of a timing technique that compromises responsiveness and flow preservation. They established four requirements for activation that are complementary to one another:
\begin{itemize}
  \item \emph{Immediate alerts} (5 groups): To avoid cascading failures and unnecessary effort, alerts should be triggered immediately once syntax or visibly critical errors occur. This immediateness is fine as long as the signal is not too large and does not obscure the editor. This will allow for a speedy acknowledgment without distracting from the task that is currently being performed (\textit{``...while I'm typing, but non-blocking''} - G5).
  
  \item \emph{On-demand activation} (6 groups): Several groups emphasized manual invocation, such as shortcuts and command palettes, as the default method for noncritical inspections, to maintain the autonomy of developers and to prevent any needless disruptions while they were ``in the zone'' (\textit{``I am annoyed by automatic alerts''} - G6). When conducting an investigation on the code or performing fine-grained editing, manual scans are considered highly effective.

  \item \emph{Post-writing triggers} (4 groups): It was suggested that batching alerts at natural pauses, such as after finishing a code chunk, on save, pre-commit, or when running tests, was a good compromise that would decrease distraction while maintaining a feedback that was timely enough to direct the next step (\textit{``...when I finish an IF block and I start coding something else''} - G3).
  
  \item \emph{Context-aware timing} (3 groups): The timing should be adjusted according to the developer's actions (such as typing versus navigating, running versus debugging), as well as the severity of the issue (\textit{``The timing of the issue should depend on the severity of the bug''} - G2). This can be accomplished by implementing a ``do not disturb'' feature or introducing intelligent throttling during intense typing to reduce alert fatigue (\textit{``I'd like the AI to deliver automatically in the relevant moments''} - G1).
\end{itemize}

\paragraph{How should alerts be customized?}
The user's expertise, the project rules, and an adaptive behavior that has been learned through interaction are all included in the customization preferences for alerting:
\begin{itemize}
  \item \emph{Experience-based adaptation} (6 groups): Unlike experts, novices gain value from more detailed text, examples, and linkages, while experts prefer concise signals and immediate access to proof. As a result, verbosity should be appropriately adjusted depending on the developer's profile (``\textit{...an expert developer may not need minor alerts. Also, the verbosity of the alert should be adjusted accordingly} - G5).
  \item \emph{User-controlled customization} (4 groups): The options for auto-run versus manual scans, severity levels, snooze rules, and noise filters should be provided based on each project (\textit{``I'd leave to the user the choice about timing''} - G1). Additionally, specific patterns should be included in the ``don't show this again'' option. 
  \item \emph{Contextual adaptation} (4 groups): Different thresholds and gating criteria should be implemented as the code advances closer to release, or depending on the application context, to ensure that alerts are aligned with repository norms and the development phase (``...personalize the sensitivity depending on the domain or criticality of the software'' - G6).
  \item \emph{AI inference and automation} (3 groups): Use the actions of accepting and dismissing alerts to learn how to modify salience and timing automatically, but always allowing users to override their preferences, guaranteeing that they are always in control (\textit{``Since the system can learn through time, it could become more proactive for recurring errors, or even predict them''} - G2).
\end{itemize}

\subsection{Explanations of Bug Alerts (Task~A, Scenario~2)}
The developers' expectations for the explanations accompanying bug notifications are reported in this section. 

\paragraph{What aspects of the bugs should be explained?}
To get from diagnosis to action, participants want explanations to include unambiguous claims of impact and origins of the bugs:
\begin{itemize}
  \item \emph{Understanding the bug’s nature} (5 groups): For the purpose of establishing a mental picture of the reasons why the alert was triggered, a brief summary of the root cause and error type (for example, out of boundary loops, null reference, or corrupted input) should be provided (\textit{``The type of the error, the error code, and a description''} - G2).
  
  \item \emph{Impact awareness} (6 groups): To facilitate prioritization in the case the developer is under time pressure, the system should include a clear articulation of concrete consequences, such as crash risk, security exposure, or performance regression (\textit{``The cause of the bug and the impact, with a description of the execution flow''} - G6).
  
  \item \emph{Practical resolution support} (3 groups): As mentioned for the alerts, it would be preferable to propose one or two targeted next steps, such as minimal viable edits, a specific test to run, or a condition to inspect, instead of providing opaque automatic fixes (\textit{"Don’t just tell me what’s wrong, show me one or two concrete things I can try right away”} - G4).
  
  \item \emph{Documentation and references} (4 groups): It can be helpful to include references to project rules, lint policies, API documentation, or small examples that are relevant to the recommendation and help with quick verification (\textit{``...by inserting links to external sources that show how to solve the bug or a detailed explanation of the problem''} - G3).
  
  \item \emph{Transparency of AI reasoning} (3 groups): An explanation of the chain of evidence (for example, the most important variables, or the stack frame), as well as a brief indication of whether or not the system is confident or uncertain about its output (\textit{``...I'd like explanations under a threshold of AI confidence not to be shown by default''} - G1).
\end{itemize}

\paragraph{How should explanations be displayed?}
The appearance of explanations should be designed in such a way that glanceability is balanced with depth on demand:
\begin{itemize}
  \item \emph{Inline and embedded options} (5 groups): It is ideal for quick checks to have a short tooltip or pop-up that is close to the code line and provides the rationale in a single sentence. An embedded clickable "More" feature would allow the user to open the full explanation (\textit{``...a pop-up at the error position with a summarized explanation, and a button to request further details''} - G3, \textit{``A mixed modality, a pop-up with an initial summary and a lateral panel for details''} - G6).
  \item \emph{Expandable and separated views} (4 groups): When it comes to longer narratives, call traces, code \emph{diffs}, or code snippets that demonstrate the fix, it is preferable to incorporate an expandable sidebar or a separate panel (\textit{``I'd like a separate view for explanations [with respect to the coding area]''} - G5).
  \item \emph{Interactive presentation} (4 groups): There were some groups that advocated for interactive exploration, such as a chatbot or a step-by-step walkthrough, with the goal of asking ``why this is happening and not that?'', or to evaluate and consider various alternative solutions (\textit{``...a side panel, with a chatbot integrated''} - G4).
  \item \emph{Auxiliary modalities} (2 groups): Clarifying logic without overloading the main text can be accomplished with the help of optional visual aids such as simple charts and minimal diagrams, which were considered optimal for complex control and data flows (\textit{“Sometimes a small diagram or chart would explain it faster than a long paragraph”} - G6).
\end{itemize}

\paragraph{Where should explanations be displayed?}
Participants, as previously mentioned, supported the idea that the explanation must be visible without diverting attention, incorporating a primary view that integrates more detailed explanations:
\begin{itemize}
  \item \emph{Lateral panels and sidebars} (6 groups): All the groups agreed on the idea that the primary location for detailed reasoning, traces, examples, and linked resources should be a sidebar that can be docked and expanded, or a bottom panel when the IDE allows it, making it simple to find explanations while maintaining the main coding editor clean \textit{(``A sidebar on the right which I can open during all the phases of the development process and close when I don't need it'' }- G4).
  \item \emph{Integrated code views} (3 groups): A lightweight inline view that is anchored to the implicated code (for example, a peek below the line or a focused highlight) allows to connect the explanation directly to the code, which helps make sense of the situation relatively quickly (\textit{``Near the implicated code in case of severe errors''} - G3).
  \item \emph{Non-invasive positioning} (4 groups): When tooltips or small expandable popovers are not invasive and do not obscure selections, completion menus, or run and debug controls, they are considered to be acceptable (\textit{``Pop-ups should adapt considering the free space in the editor in order to minimize overlappings with the code''} - G1).
  \item \emph{Dynamic and hybrid placement} (2 groups): When the explanation becomes more extensive or when the editor becomes messy, some groups proposed that the system should automatically transition from a near code preview to the sidebar (\textit{“If the explanation gets too long, it should move to a panel automatically so it doesn’t overlap the code”} - G4). 
\end{itemize}

\paragraph{When should explanations be displayed?}
According to the preferences about timing, there is a significant predisposition for progressive disclosure, with some exceptions made for high-risk situations:
\begin{itemize}
  \item \emph{On-demand delivery} (6 groups): a predominant preference suggests that explanations should be collapsed by default and only exposed when the developer requests it, for example, through clicks, hovers, or shortcuts. With this approach, the coding flow is preserved, and the cognitive stress caused by unexpected or unwanted content is reduced (\textit{``[Explanations] only on demand, I don't want to see them if I don't need them''} - G3).
  \item \emph{Automatic for critical bugs} (4 groups): As anticipated in previous discussions, automatic expansion is appropriate for high-severity or safety-critical findings (including evident syntax problems), when fast comprehension will prevent costly mistakes (``\textit{Depending on the bug's severity, proactively if the bug is impactful, on demand for minor severity}'' - G4). 
  \item \emph{Post-interaction timing} (2 groups): Similarly to the previous task, in addition to the alert moment, some groups encouraged revealing explanations following a save or commit or alongside test runs, which are times when developers naturally pause and are most likely ready to focus on details and errors in the code (\textit{“I’d prefer explanations to pop up after I save or run the code, when I actually have time to read them” }- G5).
\end{itemize}

\paragraph{How should the explanation be customized?}
The participants demanded explanations that were appropriate in size and that took into account their preferences, skills, and the context:
\begin{itemize}
  \item \emph{Experience-based detail levels} (6 groups): The system should compose explanations in a more discursive and clear way, as well as provide brief conceptual reminders when useful, in the case the user is a novice developer (\textit{``...less technicisms for novices''} - G4). Experts, on the other hand, prefer language that is more technical and that directly links to evidence (\textit{``The explanation must demonstrate to expert users that what is claimed is reliable'' }- G1). Users should be able to control the verbosity, therefore, by a minimal "profile" option setting or a quick toggle.
  \item \emph{User-controlled preferences} (4 groups): Developers would be able to bring the system into alignment with their workflow if allowed to control the detail level, and explanation timing (\textit{``The sensitivity of the AI should be customizable''} - G2).
  \item \emph{Contextual adaptation} (4 groups): The explanations should include reference to the norms of the repository, adjust to the type of file, and consider historical trends in order to make the recommendations less generic and more project-specific (\textit{``...on the basis of the system's knowledge of the project''} - G6).
  \item \emph{AI-driven personalization} (2 groups): Again, similarly to the previous task, the system may learn from follow-up queries to modify future depth and timing, while always preserving explicit user overrides (\textit{``I'd like the AI to have memory of my interaction with it [chatbot] and infer my preferences and needs''} - G1).
\end{itemize}

\subsection{Interactions with Code Readability Indicators (Task~B, Scenario~1)}
\begin{figure}[ht]
    \centering
    \includegraphics[width=\linewidth, trim=0 4cm 0 0, clip]{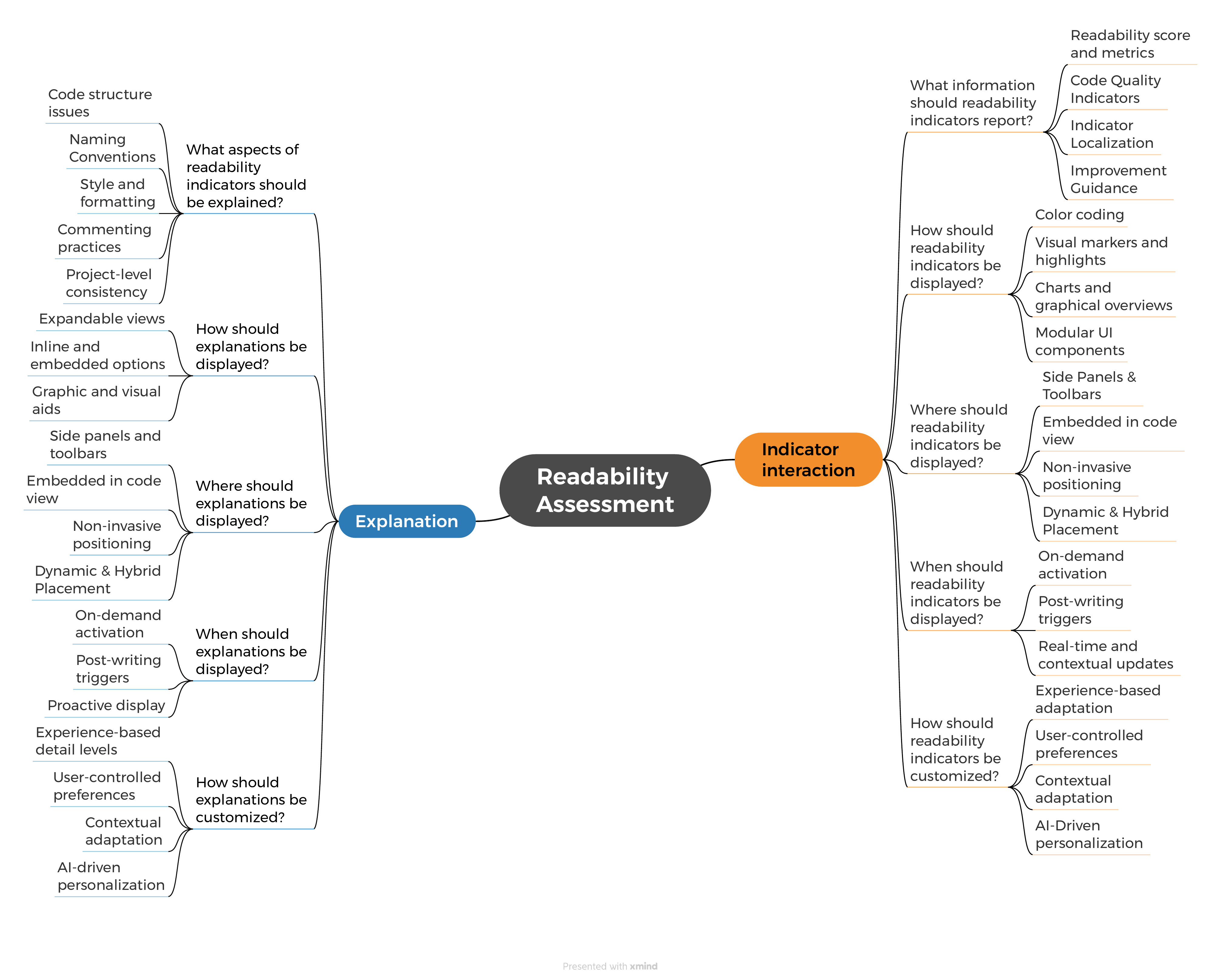}
    \caption{Themes emerged by analyzing the developers' answers to the questions on the code readability assessment task.}
    \label{fig:findings-ui-readability}
       
\end{figure}

In this subsection, we report the themes that have emerged regarding developers' expectations concerning the readability indicators that appear within the IDE. As done in previous sections, the findings are categorized according to the questions that direct the conversations throughout the workshops. Every single finding can be traced back to a single topic that arose during our investigation. An overall view of the results is depicted in Figure \ref{fig:findings-ui-readability}.

\paragraph{What information should readability indicators report?}
Developers asked for traceable, localized, and actionable content:
\begin{itemize}
  \item \emph{Readability score and metrics} (5 groups): A numerical or percentage-based objective score that includes hierarchical granularity (file, class, function, or line) and temporal evolution, which also enables comparisons between the state of things before and after (\textit{``...a contextually aware readability score''}- G4).
  \item \emph{Code Quality Indicators} (5 groups): Signals at the factor level (name, indentation, formatting, structure, comments) that help to justify why a line or block decreases the understandability of the code, rather than simply stating that it is difficult to read (\textit{``I'd like to have information about the different readability metrics separately''} - G2).
  \item \emph{Indicator Localization} (4 groups): It would be useful to provide a feature that enables adjustments that are tiny and well-targeted if there are clear pointers to the particular lines or spans (including multi-line sections) that reduce the score (``\textit{The interface should contain readability indicators linked to part of code}'' - G6).
  \item \emph{Improvement Guidance} (4 groups): Specific micro suggestions (e.g., renaming, extracting, splitting, reflowing) that are connected to the highlight, together with a brief preview and a remark on the expected outcome, to make the user aware and guide their expectations (\textit{“Give me tiny tips like ‘rename this’ or ‘split this function’”} - G1)
\end{itemize}

\paragraph{How should readability indicators be displayed?}
According to the developers, alerts should be shown through:
\begin{itemize}
  \item \emph{Color coding} (4 groups): To display status and severity at a glance without interfering with syntactic colors, encodings that are gentle and easy to read (with sufficient contrast) resulted preferable (\textit{``...differentiating colors based on the type of the issue''} - G3).
  \item \emph{Visual markers and highlights} (4 groups): Results can be linked to specific lines using underlining, gutter markers, or darkened spans (\textit{``Underlining or highlighting the low readability code spans, along with a fix suggestion that appears on hover so that intrusiveness is kept minimal''} - G1). This prevents the results from becoming confused with other lint or error decorations.
  \item \emph{Charts and graphical overviews} (3 groups): Small heatmaps and factor bars are examples of compact graphics that could provide a summary of the current status of the page or file and illustrate trends over time to demonstrate whether or not local adjustments have an effect (\textit{``By clicking on the readability score a graph could appear showing where the readability is lower''} - G5).
  \item \emph{Modular UI components} (3 groups): Widgets that are dedicated to aggregating concerns, such as a "Readability" panel, would support filtering by factor and eventually collapse when they are not required, thereby reducing the amount of distractions (\textit{``The tool should offer an optional view with the information about readability, aggregated by their type''} - G6).
\end{itemize}

\paragraph{Where should readability indicators be displayed?}
Concerning the positioning of the indicators, participants favored a localization plus overview pattern:
\begin{itemize}
  \item \emph{Side Panels \& Toolbars} (5 groups): The preferred location for information such as breakdowns, trend lines, and navigation between occurrences was a collapsible panel, which should also remember the layout state for each project (\textit{``...in a separate panel, maybe at the bottom or on the right with respect to the editor''} - G4).
  \item \emph{Embedded in code view} (3 groups): Cues that are close to the code, such as inline badges and hover hints, indicate the specific spans that require attention to reduce eye travel and context switching (\textit{``...in a row below the code, similar to completions''} - G3) .
  \item \emph{Non-invasive positioning} (3 groups): It is essential that tooltips and popovers do not obstruct selections or completion options and that they are accessible via the keyboard and may be dismissed rapidly (\textit{"Pop-ups are fine as long as they don’t block the code I’m trying to read”} - G6).
  \item \emph{Dynamic \& Hybrid Placement} (2 groups): A solution to avoid the coding area from becoming overcrowded could be represented by folding inline text into the panel on smaller screens or when numerous flags/alerts are open (\textit{“If there are too many indicators inline, push them to a panel so my editor doesn’t get messy”} - G2). 
\end{itemize}

\paragraph{When should readability indicators be displayed?}
Participants favored timelines that preserve momentum while maintaining guidance that is quite near to the point at which improvement is expected:
\begin{itemize}

  \item \emph{On-demand activation} (6 groups): It is essential to avoid noise throughout the coding process, and this could be achieved by allowing users to run a readability assessment through a shortcut or command when they are ready to refactor and dedicate to readability issues (\textit{``Manual activation with a button'' }- G1). 
  \item \emph{Post-writing triggers} (4 groups): During natural pauses, such as when a save is being made, before a committing changes to a repository, or after tests, the indicators could be provided so that suggestions can be received when the cognitive load is reduced and attention can be shifted to clean up the code and make it more readable for others (\textit{``...after a working test run with no errors''} - G5, \textit{``Before I am pushing to GitHub''} - G3).
  \item \emph{Real-time and contextual updates} (3 groups): Allow small changes to be made in small steps (in moments such as sitting down after typing). The system could implement an intelligent indicator throttling mechanism in order to reduce distraction (\textit{``When the AI understands I am done with coding''} - G2). 
\end{itemize}

\paragraph{How should readability indicators be customized?}
Expertise, project norms, and preferences that have been learned through previous interactions are all included in customization cues:
\begin{itemize}
  \item \emph{Experience-based adaptation} (3 groups): Short messages that are aimed at professionals/experts; a little more educational indicator with a few examples for beginners (\textit{``Less text depending on the developer's experience level''} - G3). Users could be allowed to switch between the two with a simple toggle.
  \item \emph{User-controlled preferences} (5 groups): The ability to choose between auto and manual analysis, readability factors to dismiss, strictness thresholds, and visualization density on a per-project basis allows teams to adjust the amount of noise they experience (\textit{``I'd like to customize the visibility of the indicator''} - G1, \textit{``I'd let the user choose the timing of the indicator''} - G6).
  \item \emph{Contextual adaptation} (4 groups): It is important to connect the readability alerts with the style guides and domain conventions of the repository. For example, large lines are allowed in data tables but not in business logic (\textit{``The AI should adapt on the basis of the project''} - G2, \textit{``The AI understands how to personalize the indicator depending on the specific project''} - G5).
  \item \emph{AI-Driven personalization} (3 groups): Through the use of transparent and easily reversible overrides, the system could learn from accept/dismiss patterns to eliminate repetitive noise and promote commonly addressed issues (\textit{“If I keep ignoring a kind of suggestion, the tool should show it less and focus on what I actually fix”} - G1).
\end{itemize}

\subsection{Explanations of the Readability Assessment (Task~B, Scenario~2)}
This subsection describes how explanations should justify the score and guide improvement. 

\paragraph{What aspects of readability indicators should be explained?}
Explanations should make the readability score and indicators understandable, and guide the user to implement improvements by mitigating the causes that lower the score:
\begin{itemize}
  \item \emph{Code structure issues} (4 groups): Long lines or functions, extensive nesting, or multibranch complexity were pointed out as primary factors to be explained, since these kinds of issues are the least immediate to understand and address ("\textit{The tool should explain on 'spaghetti' code lines}" - G2).
  \item \emph{Naming Conventions} (5 groups): An explanation of why naming is unclear or inconsistent would be helpful, with compact examples of better names that match project patterns or common norms (\textit{"...how much naming of functions or variables adapts to the overall code, with motivations"} - G1, \textit{"Priority on naming conventions"} - G4).
  \item \emph{Style and formatting} (5 groups): Developers want indentation, spacing, and formatting choices that hinder readability to be explained, paired with tiny ``before$\rightarrow$after'' diffs to illustrate the fix (\textit{"Report the motivations behind factors that lower the score and examples on how to improve it"} - G5).
  \item \emph{Commenting practices} (3 groups): Missing or obsolete comments can significantly affect code clarity and, especially in the second case, it is hard for a developer to identify comments that need refactoring, that are insufficient (\textit{"Suggest more comments for hard to read blocks that prioritize performance"} - G1), or that are no longer required. 
  \item \emph{Project-level consistency} (4 groups): Explanations should encourage alignment with repository style guides and team habits (\textit{"They should refer to project conventions or documentation"} - G2), eventually clarifying what is a contextual hard rule to follow versus a default readability good practice.
\end{itemize}

\paragraph{How should explanations be displayed?}
The form in which explanations are presented should  balance between glanceability and depth:
\begin{itemize}
  \item \emph{Expandable views} (5 groups): The most prevalent approach consists of a richer panel ranking readability factors by contribution and linking to affected lines (\textit{"An approach with highlighted lines linked to an external panel"} - G6). This could also be integrated with optional previews of the expected score change after a proposed edit.
  \item \emph{Inline and embedded options} (2 groups): It was suggested by a few groups that a concise explanation on hover/peek, which would be related to the code highlight, should be included. This explanation would answer the question ``why this line?'' in an easy-to-understand way ("\textit{A small pop-up with minimal explanation, expandable for further details}" - G4).
  \item \emph{Graphic and visual aids} (4 groups): Compact visuals (such as a factor legend, or a small concept map) would help the developer in assessing the overall readability of the code at a glance and choose the best action to take next (\textit{"With a legend of readability issues grouped by type, maybe using different colors for different categories"} - G1).
\end{itemize}

\paragraph{Where should explanations be displayed?}
Participants asked for explanations that are easy to find, while they should never compete with the editor and the already familiar features in it:
\begin{itemize}
  \item \emph{Side panels and toolbars} (5 groups): Many preferred to have a sidebar as the ``explain it to me'' space \textit{("List in a tab on the side with expandable elements and quick fixes"} - G1). This could be used to provide the why behind the indicator in plain language, also showing a summary of contributing factors (e.g., ``cyclomatic complexity  42\%, naming 23\%, duplicate code  12\%''), and including jump links that take the developer straight to each occurrence in the file. 
  \item \emph{Embedded in code view} (3 groups): Hints placed near the code would easily connect the explanation directly to the flagged span of code. They could provide a quick ``why'' without leaving the line and, consequently, reduce eye travel (\textit{"Pop-up near the code"} - G5, \textit{"In 'shadow' at the interested code lines"} - G4).
  \item \emph{Non-invasive positioning} (2 groups): Compact overlays could be implemented, but they must avoid occluding code or interfering with editing controls. They should collapse automatically after the user takes action (\textit{“Keep the explanation small and out of the way, and hide it as soon as I start typing”} - G2).
  \item \emph{Dynamic and hybrid placement} (2 groups): Some groups mentioned an automatic feature to shift between inline peeks and the sidebar as the explanation length or density increases or decreases (\textit{"A tooltip for short info, with an additional window for longer details"} - G3).
\end{itemize}

\paragraph{When should explanations be displayed?}
Timing preferences encouraged a progressive and user-controlled disclosure of the explanations:
\begin{itemize}
  \item \emph{On-demand activation} (6 groups): All groups expected explanations to remain collapsed until they are explicitly requested (through click, hover, shortcut), to keep the focus on writing until the developer chooses to review the clarity of the code (\textit{"On user demand"} - G1, \textit{"I'd make it only on request"} - G2).
  \item \emph{Post-writing triggers} (4 groups): It was noted that in most cases, developers prefer inspecting readability explanations during coding pauses, such as after finishing a function, on a save or run, or before committing to a shared repository (\textit{"During the committing phase" }- G4, \textit{"At the end of the coding process"} - G1).
  \item \emph{Proactive display} (2 groups): Some imagined an automatic display of explanation only in rare cases, such as in the presence of significant readability score drops or violations of important team-agreed criteria, which require immediate attention (\textit{“If the readability score suddenly drops or a team rule is violated, show the explanation right away so I don’t miss it”}- G4)
\end{itemize}

\paragraph{How should explanations be customized?}
Participants demanded explanations that are appropriate in verbosity and that take into account both individuals and the context:
\begin{itemize}
  \item \emph{Experience-based detail levels} (3 groups): The system should deliver an explanation that is concise and technical for the expert developers; the prose should be more comprehensive, understandable, and contain fewer reminders for beginners ("\textit{Less expert users should receive longer and more detailed explanations}" - G1).  
  \item \emph{User-controlled preferences} (4 groups): Most groups noted that such a tool should allow the user to select both the degree of verbosity and the technical level of the writing in explanations, in order to fit the developer's preferences best (\textit{"I'd like to customize the detail level and technicality degree of the explanation"} - G5).
  \item \emph{Contextual adaptation} (4 groups): Explanations should adjust to the current programming language and the application domain. For example, what counts as ``readable'' in Python may differ from Java, and indicators should reflect that (\textit{"The explanation should be shaped around the application context, it shouldn't be generic"} - G2). Similarly, guidance needs to adapt to the role of the code. For instance, API modules require clarity at the interface, while data pipelines may need to prioritize structure and traceability. 
  \item \emph{AI-driven personalization} (3 groups): In this case, too, participants imagined a system that learns and adapts to individual and team preferences over time. They suggested that the AI could automatically infer these preferences by observing how users respond to explanations, whether they follow a suggested refactoring, dismiss it, or ask for further clarification (\textit{"The AI tool should automatically understand what to explain based on the interaction with it"} - G1, "\textit{If I accept many suggestions of a specific kind it should consider it"} - G4). Also, engaging with a chatbot or conversational agent could give the system richer signals about what kind of guidance is most useful to specific users or in specific contexts (\textit{"To me the AI system should always adapt based on my interaction with it through a chatbot"} - G6). 
\end{itemize}

\section{Elicited mental models}
This section illustrates the mental models we developed for bug detection and code readability assessment, based on the study results. As depicted in the metaphorical representation diagram in Figure \ref{fig:mental_models}, the bug detection tool is conceived as a bug detective. In contrast, the readability assessment tool is conceived as a quality coach. Both these "partners" collaborate with the developers to offer actionable feedback, critical warnings, and personalized suggestions, highlighting the important role of "trust" promoted through explanation, user control, and adaptive timing.

\subsection{Mental Model for Bug Detection}
This section illustrates the mental model for bug detection that emerged from the workshops, which sees the AI as a \textit{bug detective} that integrates into the development flow. In line with our qualitative protocol, we articulate this mental model along the same interaction dimensions used to guide the elicitation.

\subsubsection{Activation and Control (RQ1)}
Participants described activation as something that should consider both context and risk. In their view, the system should remain quiet until it matters, for example, when a safety-critical bug is introduced, which may warrant immediate correction. Minor issues can be addressed during natural pauses, such as saving, running tests, or committing changes. Nevertheless, a toggle to manually disable the alerting functionality is always recommended. Developers also envision a hybrid mode: lightweight checks running silently in the background, with manual triggers such as a check button or a shortcut, allowing them to explicitly set the tool's proactivity when needed. 

\subsubsection{Visualization and Placement (RQ1)}
Regarding how alerts should be displayed in the IDEs, participants envisioned a balance between precision and overview. Inline markers, such as highlights or badges, should link issues directly to the relevant lines, making it easy to fix problems without breaking the focus. At the same time, a collapsible side panel can serve as the "hub" for bug findings, where issues are collected, categorized, and linked back to the code. The guiding principle is "not to be invasive", and this is achieved by including compact cues near the code for new or urgent issues, but leaving fuller explanations aside in expandable panels. Finally, the interface should adapt to the developer's workspace (for example, folding inline details into the panel when the editor is already crowded) so that the coding view remains in the center.

\subsubsection{Content and Granularity of Alerts (RQ1)}
Participants conceived the alert as a structured message that has to report: i) what is wrong, ii) why it happened, and iii) what to do next. The alert has to indicate the position of the bug in the code, and a short causal explanation followed by next steps such as ``apply this edit,'' "run this test" or "check this variable". Trust is related to the tool transparency, obtained, for example, by using feedback like a severity label, a confidence indicator, or a short note about where the finding came from (a rule, a model, or a failing test). Therefore, granularity should follow a progressive pattern, with short headlines inline and optional expansion for details like traces or examples.

\subsubsection{Preferences for Explanation Delivery (RQ2)}
Explanations should be available close to the code, but never expanded automatically unless there are critical findings. Participants asked for brief rationales, such as tooltips or peeks, attached to flagged portions of the code, which could unfold into richer explanations in a side panel. The latter might include short traces, snapshots of variable values, or before-and-after code snippets. These deeper explanations should also clarify how the system arrived at its conclusion (technical transparency) and which mechanism triggered the alert (operational transparency). 

\subsubsection{Personalization and Design Needs (RQ3)}
Participants emphasized that the details and proactivity of the explanations should be tailored to both the developer and the project. Experts prefer concise messages, while novices may need more didactic guidance. Similarly, safety-critical projects require the earlier identification of issues related to exploratory coding. Personalization should be supported both statically (via profile settings or per-project policies) and dynamically (by learning from interactions, such as avoiding alerts that a developer routinely dismisses). Control mechanisms like "snooze" or "never show this again" should be explicit in scope (line, file, or repository) and reversible. However, the same finding should be deliverable in different ways (short or long, proactive or quiet) without ever affecting trust or control.

\subsection{Mental Model of Code Readability}
Regarding the mental model of code readability indicators, we can conclude that developers view AI as a code quality coach that is metric-based, contextual, and customizable. 
\label{sec:readability-mm}

\subsubsection{Activation and Control (RQ1)}
Participants described activation as something that should respect both goals and flow. In their view, readability indicators should appear when they best support improvement, not while someone is still sketching an idea. Lightweight checks can run silently in the background, displaying results at natural pauses such as after saving, finishing tests, or preparing a pull request. Developers want the option to call indicators on demand, when they believe it is the right time, for instance, during a refactoring session. Automatic indicators, on the other hand, are only expected when a particular event occurs, such as a function that becomes unusually long or when a sudden, significant drop in readability occurs. In this way, the system can help without breaking concentration. 

\subsubsection{Visualization and Placement (RQ1)}
Participants devised visualization as achieving two complementary goals: pinpointing precisely what needs attention and providing a quick overview of the overall quality. The former can be obtained with inline highlights (e.g., using underlines or gutter markers), which help act on code without leaving context. Small "factor chips" (naming, structure, comments) can help explain why a line was flagged in plain sight. The latter can be achieved instead with a collapsible sidebar that acts as the central hub, where developers can filter by factor or file, view trends over time, and quickly return to the code.
In general, the placement of the indicators should be non-intrusive, with compact cues near the code for quick checks. Richer details and comparisons are available in an expandable side panel that does not clutter the main view. The interface should also adapt to the editor space, moving inline information to the sidebar when it becomes crowded, so that the primary focus remains on the code and is not interrupted.

\subsubsection{Content and Granularity of Indicators (RQ1)}
For this task, developers envision the system as having a \textit{coaching} role, providing a transparent and progressive flow: highlighting what is hard to read, explaining why it matters, showing how to improve it, and illustrating what would change. They want an objective overall readability score broken down into levels (file, class, function, line) and factors (naming, formatting, comments, structure). Indicators should be specific and localized, pointing to concrete spans, such as "this function exceeds 50 lines," rather than vague advice. Suggestions should be actionable and small, such as renaming a variable, splitting a function, or amending a comment; minimal examples can accompany the suggestion to illustrate the better form. Participants also asked for \textit{previews} of how a refactor would affect the score and comparisons to past versions or team norms: This can help understand and quantify the expected impact of applying the suggestions. 

\subsubsection{Preferences for Explanation Delivery (RQ2)}
Similarly to the indicators, explanations should be located close to the code and expanded only when needed. Developers expect a short rationale to appear on hover or inline peek to clarify why a particular line was flagged. For more detailed explanations, the sidebar can unfold into a narrative supported by evidence, including factor breakdowns, before-and-after comparisons, and links to each occurrence. Explanations should also reflect the broader context, showing how readability has evolved in a file and whether it aligns with team standards. When the case is borderline or stylistic, participants wanted the explanation to make that nuance visible rather than framing it as a rigid violation. For impactful issues, such as modifying tens of lines of code, developers also imagined ``what-if'' previews that simulate the improvement and demonstrate the gain without actually affecting the code, helping them prioritize their efforts.

\subsubsection{Personalization and Design Needs (RQ3)}
Participants clearly expressed the need for the readability coach to adapt to people, projects, and languages. Detail levels should scale with experience, with concise text for experts, and more didactic suggestions for beginners. Preferences should be configurable (e.g., by setting profiles like ``Beginner'', ``Expert'', or ``Team'') and dynamically learned from behavior, similarly to the case of the bug detection scenario; an example would be lowering the weight of factors a developer consistently dismisses and prioritizing those they act on. Project standards also matter, as teams may want to import their style guides or templates and adjust thresholds so that accepted practices are not unfairly flagged. Language and domain rules should be adjusted according to the specific programming language, and expectations may vary across different application domains. Developers requested lightweight control options, such as snoozing or disabling specific metrics with a clear scope and easy rollback, which would enable the system to adapt to team practices rather than enforcing default norms.

\section{Design tensions}
\label{sec:design_implications}
This section synthesizes the study's findings by discussing design tensions for IDEs that support automated bug detection and code readability assessment. These points can guide the development of HCAI IDEs that improve the user interaction in both tasks. In the following, we will use "suggestion" to refer to bug detection \textit{alerts} and code readability \textit{indicators}. 

\textbf{Proactive Activation vs. Preserving Flow:} Systems should not disrupt the developers' activities; therefore, proactive interventions should be context-sensitive and aligned with workflow rhythms and risk levels \cite{pu2025assistance}. However, this balance is not trivial to achieve, as it should be understood which situations involve high-severity issues — requiring automatic activation — and which suggestions are minor and are to be deferred or batched. In any case, the developer should be allowed to set a "do-not-disturb" or "review" mode that respects their development workflow. 
    
\textbf{Concise vs. Detailed Suggestions:} AI-generated suggestions must be carefully delivered based on the suggestion's content and length. Implementing a layered design for suggestions may help achieve this by displaying cues inline — providing immediate and available suggestions — and using expandable side panels — providing detail and overview as aggregated insights, deeper explanations, and historical exploration. This layered, incremental approach can help support both rapid debugging and long-term reasoning \cite{bo2024Incremental}. 

\textbf{Actionable vs. Didactic Suggestions:} Novice and expert developers have different needs when receiving suggestions for bug fixing and code readability. Generally, feedback should be practical and provide concrete next steps to fix a bug or improve the code readability. This can be supported by providing specific and localized advice, avoiding vague and abstract suggestions, and offering small and actionable recommendations. However, novice users might benefit from suggestions that are more likely to educate them rather than receiving minimal information; this may be done by offering suggestions that are accompanied by explanations that provide general rules or describe practical examples. 
    
\textbf{Concise vs. Detailed Explanations:} Explanations are yet another element that can contribute to increasing the user's workload \cite{Mayer2001Cognitive, Kulesza2013Ways}; therefore, explanations should be delivered progressively, e.g., by accompanying suggestions initially with concise rationales, which lead to deeper causal or metric-based details. Furthermore, more detailed explanations can be provided within expandable views to include additional reasoning or examples. This progressive disclosure may strike a balance between immediacy and depth, supporting both novice and expert developers at varying levels. 
   
\textbf{Achieving Trust:} Trust is crucial for adopting an AI-based system, and designing such systems to be trustworthy requires different aspects to be addressed \cite{Lee2004Trust}. Among these, transparency and soundness are necessary to enable users to trust the AI suggestions \cite{Kulesza2013Ways, kocielnik2019Accept} and accept them appropriately, even when the system is imperfect \cite{kocielnik2019Accept}. Therefore, AI-powered IDEs should clearly indicate what an issue is and how it was identified, including its risk severity and level of confidence. Another aspect to increase trust is providing users with more control \cite{Bertrand2023Selective}. Having configurable \textit{control levers} — i.e., reversible actions (undo, dismiss), scoping options (line, file, project), and snooze functions — lets developers regulate the level and scope of assistance.
    
\textbf{Individual- vs. Team-based Personalization:} Personalization and adaptivity of the system resulted in being a critical aspect desired by most of the developers we included in the workshops. Therefore, the system's features should be configurable to individual expertise, but also be team-aware, following team norms and project needs, and aligning feedback with collective practices and policies. This can increase consistency, reduce friction, and enhance adoption at the organizational level \cite{russo2024navigating}.
    
\textbf{Differences in Role-Framing:} Defining a role at the beginning of the interaction can help shape the mental model of the developers, set the right expectations, and increase perceived usefulness and ease of use \cite{zakharov2025ai, GooglePAIR2019People, kocielnik2019Accept}. From what emerged in the workshops, bug detection should be designed and communicated as a \textit{bug detective}, while readability scoring should be framed as a \textit{quality coach}.

These design aspects are related to IDEs, but include several valuable aspects that can be generalized to any domain where AI provides evaluative judgments, such as medical diagnostics, educational support, or security monitoring.
While our main contribution is understanding and expressing developers’ mental models, we have also gained some valuable insights through early prototypes presented in Section \ref{sec:prototype}.

\section{Threats to Validity}

\subsection{Internal validity} 
Since ideas were generated and refined in group co-design discussions, outspoken participants may have disproportionately shifted consensus towards their preferences. This means some needs and design cues may reflect individual judgments rather than group dynamics.
To mitigate this threat, the conductors stimulated discussion among all participants; moreover, brief individual reflection steps allowed every participant to express their own opinions. 

In addition, six facilitators conducted workshops at three sites. Even with an effort to avoid exposing any personal preferences during the sessions, slight differences in prompting, pacing, or emphasis can influence what participants say, affecting comparability between groups. This problem has been mitigated by involving all six HCI researchers in designing a shared protocol that was followed consistently during the workshops.

Finally, each workshop covered bug detection and readability assessment within the same session. Despite having balanced task orders across sessions, exposure to concepts such as activation timing and explanation style in the first task could prime responses for the second task, possibly increasing the risk of convergence across tasks. 

\subsection{Construct validity}
Participants differed in their prior exposure to AI-assisted programming tools — e.g., code completion, Copilot-like systems. As a result, the elicited mental models may conflate different subpopulations (experienced vs.\ inexperienced with such tools), which can blur the interpretation of developers' preferences regarding proactivity, explanation depth, and control. Therefore, mental models may reflect expectations rooted in prior habits that users have with other tools, rather than tool-agnostic cognition. 

\subsection{External validity} 
The sample was drawn from three universities in Italy and included many students and professional developers; therefore, cultural, organizational, and tooling ecosystem diversity is limited. This entails that generalization to larger or more heterogeneous contexts — such as globally distributed teams, safety-critical domains, or organizations with different governance and style policies — is constrained, potentially making the observed preferences not fully transferable to those settings. 

\section{Reifying the mental model with early prototypes}
\label{sec:prototype}
This section presents early-stage prototype user interfaces for \textit{bug detection} and \textit{code readability assessment} tools, which are based on design principles distilled from the study. The goal is not to propose or evaluate complete systems. Instead, it aims to demonstrate how the identified principles and mental models can guide the design of AI-assisted bug detection and tools for assessing code readability. The prototypes serve as simple sketches to connect the theoretical insights with possible applications.

\subsection{Bug Detection Task}
\begin{figure}[!htbp] 
  \centering
  % Replace with your actual path/filename
  \includegraphics[width=1\linewidth]{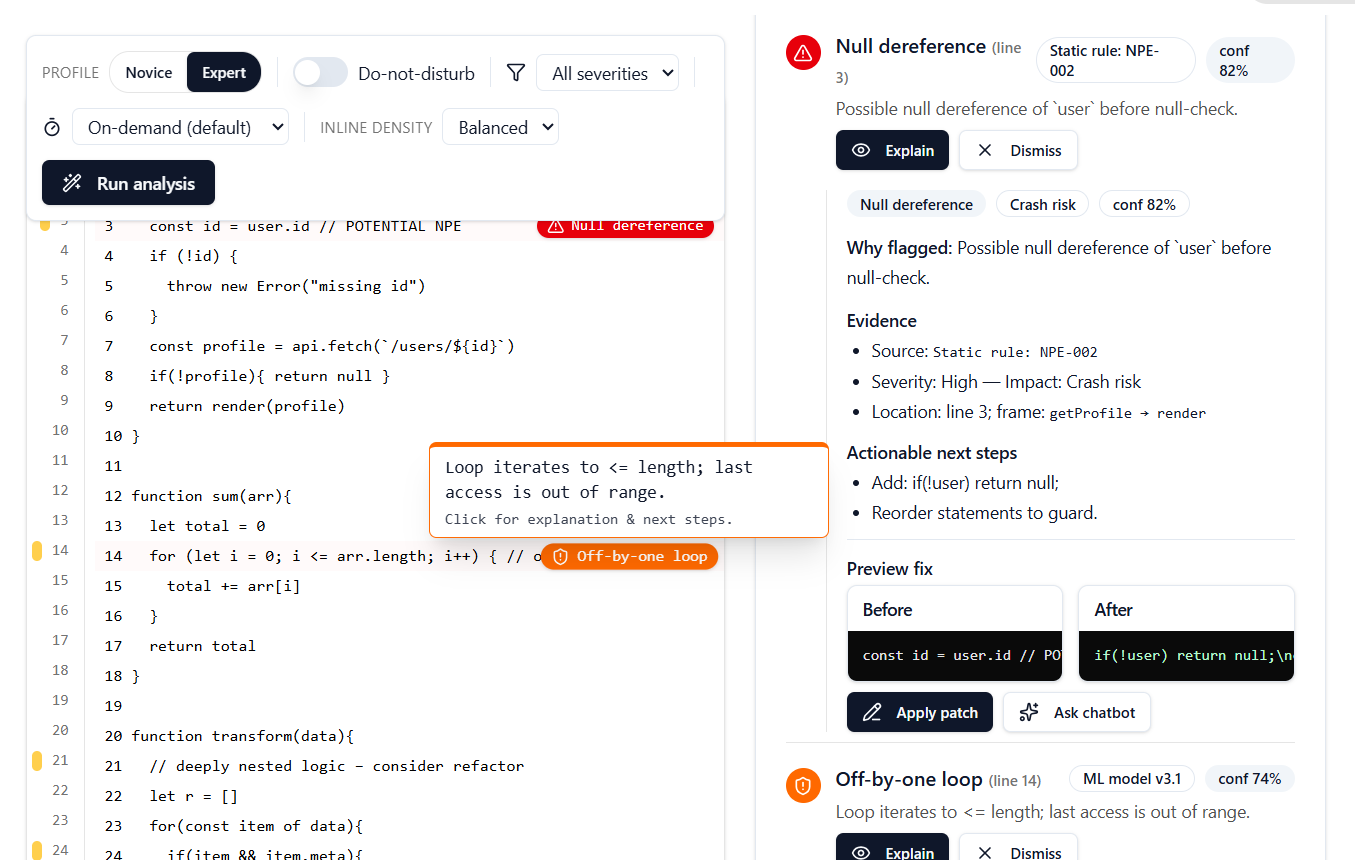}
  \caption{Full interface with the \textit{Alerts} tab selected. The first alert is expanded in place, showing rationale, evidence, confidence, and a preview patch.}
  \label{fig:proto-alerts-expanded}
\end{figure}

This tool, reported in Figure~\ref{fig:proto-alerts-expanded}, highlights potential bugs/defects directly in the editor via inline, severity-colored badges anchored to the exact line. Hovering a badge reveals a concise tooltip, while clicking \emph{Explain} expands the alert \emph{in place} within the \textit{Alerts} tab. The expanded panel provides progressive details: (i) a plain-language rationale, (ii) structured evidence (rule/model source, severity, confidence, code location), and (iii) concrete next steps with a preview of a candidate patch (before/after). Figure~\ref{fig:proto-alerts-expanded} shows the full interface with the \textit{Alerts} tab selected and the first alert expanded.

\subsection{Code Readability Task}
\begin{figure}[!htbp] 
  \centering
  % Replace with your actual path/filename
  \includegraphics[width=1\linewidth]{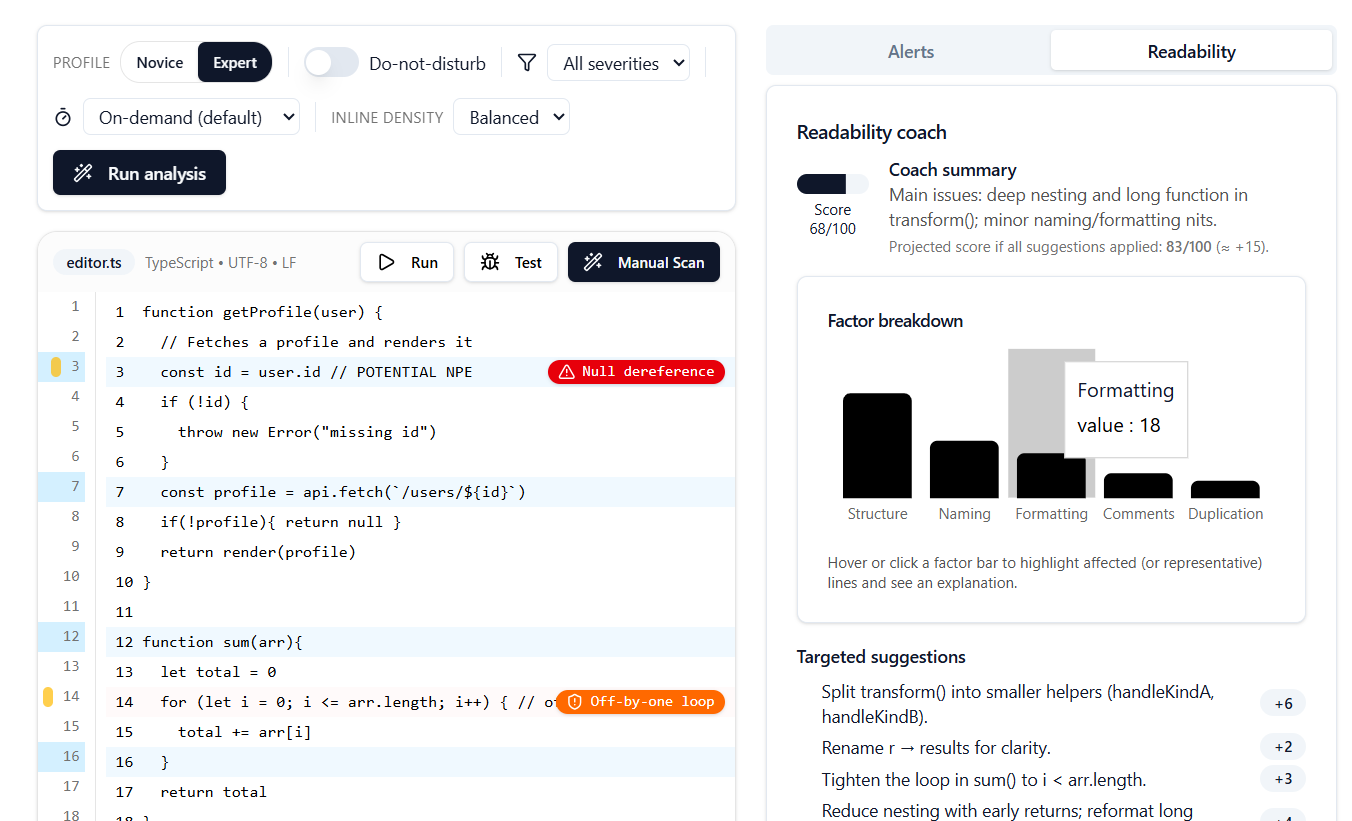}
  \caption{Full interface with the \textit{Readability} tab selected. Hovering the \textit{Formatting} factor highlights relevant (or representative) lines.}
  \label{fig:proto-readability-formatting}
\end{figure}

The \textit{Readability} tool, as shown in Figure~\ref{fig:proto-readability-formatting}, consists of a view on the right side of the IDE that presents an overall score (0–100), a factor breakdown, and targeted suggestions with estimated score improvements. When the user hovers or clicks a factor bar, the editor highlights corresponding lines to connect the metric to concrete code locations (for example, lines 3, 7, 12, and 16 in Figure~\ref{fig:proto-readability-formatting}). An explanation clarifies why the factor matters and what to change. Estimated per-suggestion deltas (e.g., “+2 points” for a variable rename) and a projected cumulative score help developers prioritize impact, aligning with personalization and adaptivity. Figure~\ref{fig:proto-readability-formatting} shows the full interface with the \textit{Readability} tab selected and a hover on \textit{Formatting}.

\section{Conclusion}

In this study, we report the results of six co-design workshops aimed at investigating the mental models of developers when interacting with IDEs with AI-augmented capabilities for bug detection and code readability assessment. 
This study contributes to ongoing discussions in HCI about human-centered AI \cite{Shneiderman2020HumanCentered} by showing how developers devise AI-assisted IDEs as \textit{augmentative} systems that should carefully balance user control, explainability, and adaptivity. 
To support effective adoption of these systems, we need \textit{context-sensitive} activation methods that minimize disruption \cite{byung2015interruption} while maintaining human initiative, progressive forms of explanation that improve transparency and trust \cite{bo2024Incremental}, and personalization mechanisms that fit both individual expertise and team practices. Additionally, how developers perceive the role of AI shapes the collaboration process \cite{zakharov2025ai}---for example, in the bug detection task, the AI is viewed as a diagnostic partner, while in readability scoring it is seen as a coaching assistant. 
Overall, the design implications that emerged from this study---e.g., about explanation depth, suggestions timing, system tailoring---may extend beyond the domain of software engineering; this can offer transferable guidance for the development of interactive AI systems that provide evaluative feedback, like medical decision support, educational technologies, and security monitoring. 

As future work, it will be necessary to investigate empirically whether the theoretical implications actually translate into HCAI tools that are useful and well-accepted by developers. With this goal, we will conduct formative usability studies on a prototype (an early version available in the additional materials) that builds on the theoretical findings emerging from this study. Future studies with developers will be needed to understand whether such a tool improves their effectiveness, efficiency, and satisfaction in AI-supported tasks like automated bug detection and code readability assessment. 

\section*{Acknowledgements}\label{sec:acknowledgements}
This research is partially funded by the Italian Ministry of University and Research (MUR) and by the European Union - NextGenerationEU, Mission 4, Component 2, Investment 1.1, under grant PRIN 2022 ``DevProDev: Profiling Software Developers for Developer-Centered Recommender Systems'' — CUP: H53D23003620006.

The research of Francesco Greco is funded by a PhD fellowship within the framework of the Italian “D.M. n. 352, April 9, 2022”- under the National Recovery and Resilience Plan, Mission 4, Component 2, Investment 3.3 - PhD Project “Investigating XAI techniques to help user defend from phishing attacks”, co-supported by “Auriga S.p.A.” (CUP H91I22000410007).

\section*{Generative AI Disclosure}
Grammarly Pro has been used for copyediting and language polishing. Gemini has been used to stylize a teaser image created by the authors.

\bibliographystyle{elsarticle-harv}
\bibliography{sample-base,exchi}

\include{appendix}

\end{document}

%% file: appendix.tex
\section{Appendix}

In this appendix, we provide additional content that enhances the transparency of the study's details and promotes replicability of the study. In particular, we report the scenarios to which users were exposed for each task, the demographic questionnaire that participants completed at the end of the study, and an excerpt from the codebook and themes developed by the two researchers during data analysis.
\subsection{Scenarios}
\label{appendix-scenarios}
In the following, we present the script used in the study to guide the elicitation of the mental model for bug detection and code readability assessment tasks.

\textbf{Bug Detection - basic.}
Sophia is a senior developer at GreenTech, responsible for maintaining a complex codebase that processes real-time sensor data to monitor urban air quality. The IDE she is using integrates an AI tool to detect bugs. During a routine update, this AI tool flags a potential bug related to sensor data parsing—a bug that, if unchecked, could lead to inaccurate pollution readings that could affect public safety decisions. Reflect on how this alert should be designed to support a senior developer’s workflow and technical insight, and then answer a set of questions we are going to ask.

\textbf{Bug Detection - explanations.}
In the same context as the previous scenario of Sophia, the IDE’s AI tool also provides an explanation outlining why the bug may occur. Consider how this explanation might be integrated into the workflow to enhance understanding and trust without overwhelming the developer, and then answer a set of questions we are going to ask.

\textbf{Code Readability - basic.}
Luca is a junior developer at FarmSoftHouse, developing a mobile app for remote smart farm monitoring. His IDE features an AI tool that automatically assesses code readability—providing a score based on structure, naming conventions, and documentation quality. Reflect on how this feedback should be integrated into the workflow to enhance understanding and trust without overwhelming the developer, and then answer a set of questions we are going to ask.

\textbf{Code Readability - explanations.}
In the same context as the previous scenario of Luca, the AI tool also provides an explanation along with the readability score. This explanation details which factors contributed to the score and offers specific suggestions for improvement. Consider how this information should be presented to balance clarity with ease of use, and then answer a set of questions we are going to ask.

\subsection{Demographic Questionnaire}
\label{appendix-questionnaire}
\begin{table}[H]
\centering
\renewcommand{\arraystretch}{1.3}
\begin{tabularx}{\linewidth}{|c|X|>{\raggedright\arraybackslash}m{4.2cm}|}
\hline
\textbf{ID} & \textbf{Question} & \textbf{Type} \\
\hline
1 & What is your age? & Number \\
\hline
2 & What is your gender? & Multiple choice: Male / Female / Non-binary / Other / Prefer not to say \\
\hline
3 & What is your academic qualification? & Multiple choice: Bachelor / Master / PhD \\
%%\hline
%4 & Self-assessed programming proficiency & Multiple choice: Low / Medium / High \\
\hline
5 & How many years of programming experience do you have? & Number \\
\hline
6 & Have you used AI-based code completion tools? & Multiple choice: Yes / No, but I know what they are and how they may work / No \\
\hline
7 & If yes to question 6: how frequently do you use such tools? & Likert 1–5 (1 = Rarely, 5 = Every day) \\
\hline
\end{tabularx}
\caption{Questionnaire items and response types}
\end{table}

\subsection{Codes and Themes}
\label{appendix-codebook}
\begin{tabularx}{\linewidth}{|>{\raggedright\arraybackslash}p{0.8cm}|
                                >{\raggedright\arraybackslash}p{0.8cm}|
                                >{\raggedright\arraybackslash}p{1.2cm}|
                                >{\raggedright\arraybackslash}X|
                                >{\raggedright\arraybackslash}X|}
\hline
\textbf{Task} & \textbf{Sc.} & \textbf{QID} & \textbf{Theme} & \textbf{Codes (frequency)} \\
\hline
\endfirsthead

\hline
\textbf{Sc.} & \textbf{Task} & \textbf{QID} & \textbf{Theme} & \textbf{Codes (frequency)} \\
\hline
\endhead

\hline
\endfoot

\hline
\endlastfoot

A & 1 & Q1 & Bug Identification And Classification & Type error (6), Classification of the error, Type of feedback (1), Filter for category (1) \\
\hline
A & 1 & Q1 & Severity And Impact Assessment & Severity error (6), Consequences/risks (4), Impact (6), runtime behavior (1), timestamp (1) \\
\hline
A & 1 & Q1 & Precise Location And Contextual Details & Bug Exact Position (6), Data context (variables, comparison values) (3), Expected values (6), Elements of tracing (3), Code span pointing (2) \\
\hline
A & 1 & Q1 & Root Cause And Diagnostic Explanation & Cause of the error (5), Explanation (4), Explanation bug flow (1), Reasons  (3) \\
\hline
A & 1 & Q1 & Actionable Feedback And Fix Suggestions & Fix suggestions (4), Possible solutions (4), Suggestions for solution (6), Web search (1) \\
\hline
A & 1 & Q1 & Source Attribution And Transparency & Source for generating the alert (2), Source of the alert (1), AI reliability metric (1) \\
\hline
A & 1 & Q2 & Color Coding & Color coding (6), Colors for severity (6), Colors for type (2) \\
\hline
A & 1 & Q2 & Iconography And Visual Markers & Icons (5), Icons for severity (1), Icons for type (1), Icons for occurrences of the bug (1) \\
\hline
A & 1 & Q2 & Evidence-Based Highlighting & Highlighting code (5), Highlighting the code with colors (4), Reference to the code (2) \\
\hline
A & 1 & Q2 & Dynamic Presentation & Animations (3), Animations of danger/alert (5), Interactive walkthrough (1) \\
\hline
A & 1 & Q2 & Modalities And Layouts & Structured layout (4), Descriptive panel (3), Chatbot (2), Audio/video (2), Grid (1), Text (and.g. \% issue) (1) \\
\hline
A & 1 & Q3 & In-Line Placement & Inline (6), Integrated in the line with visual indicators (1) \\
\hline
A & 1 & Q3 & Side Panels And Dedicated Views & Side panel (5), Separate panel (4), Sidebar expandable (4) \\
\hline
A & 1 & Q3 & Non-Invasive Positioning & Non-invasive position  (5), Not overlapping with the code (2) \\
\hline
A & 1 & Q3 & Dynamic Adaptation & Dynamic adaptation (4), Near to the focus of the user (1) \\
\hline
A & 1 & Q4 & Immediate Alerts & Immediately for critical bugs (5), Immediately for syntax errors (1) \\
\hline
A & 1 & Q4 & On-Demand Activation & On request (6) \\
\hline
A & 1 & Q4 & Post-Writing Triggers & When finished writing block (4), At commit (1) \\
\hline
A & 1 & Q4 & Context-Aware Timing & Timing inferred by the AI (3), "Do not disturb" (1) \\
\hline
A & 1 & Q5 & Experience-Based Adaptation & Detail adaptive (experience user) (6), More text for novices (3), Less details for experts (1) \\
\hline
A & 1 & Q5 & User-Controlled Customization & Personalized options  (4), "Don't show this again" button \\
\hline
A & 1 & Q5 & Contextual Adaptation & Adapt on context (3), On the basis of past interactions (4) \\
\hline
A & 1 & Q5 & AI Inference And Automation & Inferred automatically by the AI (3) \\
\hline
A & 2 & Q6 & Understanding the Bug's Nature & The root cause of the error (6), Cause of the bug (6) \\
\hline
A & 2 & Q6 & Impact Awareness & Potential impact  (5), Impact of the bug (6), Consequences/risks (4) \\
\hline
A & 2 & Q6 & Practical Resolution Support & Suggestions fix (5), Possible solutions (4), Suggestions for solutions (6) \\
\hline
A & 2 & Q6 & Documentation And References & References to documentation (4), Links to the explanation (1) \\
\hline
A & 2 & Q6 & Transparency of AI Reasoning & Reasons behind the explanation (3), Rationale of the explanation (2), Degree of confidence of the AI (1) \\
\hline
A & 2 & Q7 & Interactive Presentation & Chatbot/interactive (4), Interactive explanation  (1), Interactive walkthrough(1) \\
\hline
A & 2 & Q7 & Inline And Embedded Options & In-line integration (4), Tooltip (6), Pop-up (6) \\
\hline
A & 2 & Q7 & Expandable And Separated Views & Expandable sidebar  (4), Separate panel (4), Separate window activated inline (1) \\
\hline
A & 2 & Q7 & Auxiliary Modalities & Audio/visual aids (2), Audio explanation (1), Interactive graph (1) \\
\hline
A & 2 & Q8 & Lateral Panels And Sidebars & Panels side/right/bottom (6), Expandable sidebar (4), Side panel (5) \\
\hline
A & 2 & Q8 & Integrated Code Views & Integrated with code (3), In line (1), On the code (2), Highlight of the code (1) \\
\hline
A & 2 & Q8 & Non-Invasive Positioning & Non-invasive position (4), Small expandable window (1), Contextual tooltips  (1) \\
\hline
A & 2 & Q8 & Dynamic And Hybrid Placement & Dynamic positioning (2), Hybrid (preview near to the code, expandable a side) (1) \\
\hline
A & 2 & Q9 & On-Demand Delivery & On request (6) \\
\hline
A & 2 & Q9 & Automatic for Critical Bugs & Automatic for critical errors (4), Immediately for syntax errors (1) \\
\hline
A & 2 & Q9 & Post-Interaction Timing & After writing (1), At the same time of the alert (1), Contextual timing (2) \\
\hline
A & 2 & Q10 & Experience-Based Detail Levels & Detail based on experience (6), Simplified explanation for everyone (1), Technical details on request (3) \\
\hline
A & 2 & Q10 & User-Controlled Preferences & Customized preferences (4), On request (4) \\
\hline
A & 2 & Q10 & Contextual Adaptation & Adaptation to the application context (3), Based on past interaction (4) \\
\hline
A & 2 & Q10 & AI-Driven Personalization & Inference of preferences (2), The system learns the  preferences (1), Interaction with chatbot (2) \\
\hline
B & 1 & Q11 & Readability Score And Metrics & Score  (5), Percentage of readability (5), Trend of the readability in time (1) \\
\hline
B & 1 & Q11 & Code Quality Indicators & Naming conventions (5), Indentation and style (4), Code-based metric (1) \\
\hline
B & 1 & Q11 & Feedback Localization & Localized suggestions  (4), Highlighting of the critical lines (3), Underlying  the critical lines (1) \\
\hline
B & 1 & Q11 & Improvement Guidance & Practical suggestions  (4), Impact of the suggestions (1), Guide on improvement for novices (1) \\
\hline
B & 1 & Q12 & Color Coding & Color coding in the code (3), Bar with colored indicators (1), Color coding based on severity (4) \\
\hline
B & 1 & Q12 & Visual Markers And Highlights & Highlighting of the critical lines (3), Dedicated icon (3), Glasses icon (1) \\
\hline
B & 1 & Q12 & Charts And Graphical Overviews & Chart (2), Heatmap (1), Interactive Chart (1), Conceptual map (1) \\
\hline
B & 1 & Q12 & Modular UI Components & Statistics section (1), Separate panel (2), Alert (2) \\
\hline
B & 1 & Q13 & Side Panels And Toolbars & Side panel (5), Toolbar (1), Separate section (5) \\
\hline
B & 1 & Q13 & Embedded Views & In the toolbar (1), On top or right (2), Near the code (2) \\
\hline
B & 1 & Q13 & Non-Invasive Positioning & Separate from the code (5), Expandable small window (1), Contextual tooltip (1) \\
\hline
B & 1 & Q13 & Dynamic And Hybrid Placement & Hybrid positioning  (1), Adaptation dynamic (2) \\
\hline
B & 1 & Q14 & On-Demand Activation & On request (5), On button click (2) \\
\hline
B & 1 & Q14 & Post-Writing Triggers & At the end of writing (4), At commit (4), At push (1) \\
\hline
B & 1 & Q14 & Real-Time And Contextual Updates & During writing (3), Depending on readability (2), Live (1) \\
\hline
B & 1 & Q15 & Experience-Based Adaptation & Short explanation for experts (1), Didactic tips for novices (2), Auto-fixes only for experts (1) \\
\hline
B & 1 & Q15 & User-Controlled Preferences & Customizable preferences (4), Choice on position (1), Choice on timing (2) \\
\hline
B & 1 & Q15 & Contextual Adaptation & Contextual adaptation (3), Depending on the codebase (2), Depending on the project (1) \\
\hline
B & 1 & Q15 & AI-Driven Personalization & Inference of the preferences (2), The system learns the preferences (1), Interacting with chatbot (2) \\
\hline
B & 2 & Q16 & Code Structure Issues & Long code lines (3), Long functions (1), Nested loops (1) \\
\hline
B & 2 & Q16 & Naming Conventions & Naming of variables/functions (4), Naming uniformity (2), Semantic coherence (1) \\
\hline
B & 2 & Q16 & Style And Formatting & Indentation (4), Writing style (2), File format (1) \\
\hline
B & 2 & Q16 & Commenting Practices & Comments in the code (2), Comments/code ratio (1) \\
\hline
B & 2 & Q16 & Project-Level Consistency & Adherence to the standards (3), Coherence with the style of the project (3) \\
\hline
B & 2 & Q17 & Inline And Embedded Options & Tooltip (1), Pop-up (2), Inline (1) \\
\hline
B & 2 & Q17 & Expandable Views & Expandable (3), Separate panel (3), List (1) \\
\hline
B & 2 & Q17 & Graphic And Visual Aids & Graph (1), Legend (3), Conceptual map (1) \\
\hline
B & 2 & Q18 & Side Panels And Toolbars & Side panel (5), Expandable section on side (3) \\
\hline
B & 2 & Q18 & Embedded in Code View & Near to the code (5), Highlight of the code (1) \\
\hline
B & 2 & Q18 & Non-Invasive Positioning & Contextual tooltip (1), Small panel (1) \\
\hline
B & 2 & Q18 & Dynamic And Hybrid Placement & Hybrid positioning (1), Dynamic adaptation (2) \\
\hline
B & 2 & Q19 & On-Demand Activation & On request (6), On hover (1) \\
\hline
B & 2 & Q19 & Post-Writing Triggers & At the end of the session (1), After writing the code (3) \\
\hline
B & 2 & Q19 & Proactive Display & Proactively (1), On commit (1) \\
\hline
B & 2 & Q20 & Experience-Based Detail Levels & Simple explanation (1), Technical details on request (3) \\
\hline
B & 2 & Q20 & User-Controlled Preferences & Detail level choice (3), Choice on language style (3) \\
\hline
B & 2 & Q20 & Contextual Adaptation & Adaptation to the context application (3), Depending on the background (2) \\
\hline
B & 2 & Q20 & AI-Driven Personalization & Interaction with chatbot (2), Inference of preferences by the AI (2) \\
\hline
\end{tabularx}